\documentclass[final,12pt]{elsarticle}
\usepackage[utf8]{inputenc}
\let\today\relax
\makeatletter
\def\ps@pprintTitle{%
    \let\@oddhead\@empty
    \let\@evenhead\@empty
    \def\@oddfoot{\footnotesize\itshape
         {} \hfill\today}%
    \let\@evenfoot\@oddfoot
    }
\makeatother
\usepackage{amssymb}
\journal{}
\usepackage{lineno}
\input{macros.sty}
\begin{document}

\begin{frontmatter}

%% Title, authors and addresses

%% use the tnoteref command within \title for footnotes;
%% use the tnotetext command for theassociated footnote;
%% use the fnref command within \author or \affiliation for footnotes;
%% use the fntext command for theassociated footnote;
%% use the corref command within \author for corresponding author footnotes;
%% use the cortext command for theassociated footnote;
%% use the ead command for the email address,
%% and the form \ead[url] for the home page:
%% \title{Title\tnoteref{label1}}
%% \tnotetext[label1]{}
%% \author{Name\corref{cor1}\fnref{label2}}
%% \ead{email address}
%% \ead[url]{home page}
%% \fntext[label2]{}
%% \cortext[cor1]{}
%% \affiliation{organization={},
%%            addressline={},
%%            city={},
%%            postcode={},
%%            state={},
%%            country={}}
%% \fntext[label3]{}

\title{A New Heuristic for Rectilinear Crossing Minimization}

%% use optional labels to link authors explicitly to addresses:
%% \author[label1,label2]{}
%% \affiliation[label1]{organization={},
%%             addressline={},
%%             city={},
%%             postcode={},
%%             state={},
%%             country={}}
%%
%% \affiliation[label2]{organization={},
%%             addressline={},
%%             city={},
%%             postcode={},
%%             state={},
%%             country={}}

\author[UCA]{François Doré}
\ead{dore@i3s.unice.fr}
\author[UCA]{Enrico Formenti}

\address[UCA]{Universit\'e C\^ote d'Azur, CNRS, I3S, France}

\begin{abstract}
%% Text of abstract
A new heuristic for rectilinear crossing minimization is
proposed. It is based on the idea of iteratively repositioning nodes after a first initial graph drawing.
The new position of a node is computed by casting \emph{rays}
from the node towards graph edges. Each ray receives a mark
and the one with the best mark determines the new position.

The heuristic has interesting performances when compared to
the best competitors which can be found in classical graph
drawing libraries like
% \emph{Tulip}~\cite{auber:hal-01654518} or
\emph{OGDF}~\footnote{Open Graph Drawing Framework (https://ogdf.uos.de)}~\cite{OGDF}.
\end{abstract}

%%Graphical abstract
% \begin{graphicalabstract}
% %\includegraphics{grabs}
% \end{graphicalabstract}

%%Research highlights
% \begin{highlights}
% \item Research highlight 1
% \item Research highlight 2
% \end{highlights}

\begin{keyword}
    Graph Drawing \sep Rectilinear Crossing Minimization \sep Algorithmic Geometry
%% keywords here, in the form: keyword \sep keyword

%% PACS codes here, in the form: \PACS code \sep code

%% MSC codes here, in the form: \MSC code \sep code
%% or \MSC[2008] code \sep code (2000 is the default)

\end{keyword}

\end{frontmatter}

% \maketitle

%% \linenumbers

%% main text
% !TEX root = ./main.tex

\section{Introduction}
Graph drawing is a living research domain with an impressive
number of publications over the years. It is difficult to
say when the domain was born and who were the very first pioneers.
However, for the questions concerning our paper, one can
surely cite the seminal paper of Tutte~\cite{tutte-1963}.
In his paper, Tutte proposed an algorithm where all vertices were consecutively placed at the barycenter of the positions of their neighbours, which mimics spring forces. Afterwards, many other algorithms, called force-based models, took over the concept to draw
graphs in a \emph{nice} (and accessorily fast) manner.
An extensive compilation of force-directed algorithms can be found in a paper of Kobourov~\cite{spring_and_force_algorithms}.
In these kind of algorithms, the idea is that a graph is assimilated to a sort of particle system in which particles are identical and electrically charged. The nodes of the graph play the role of particles
and since all particles are identically charged, they tend to repel each other by Coulomb's law. However, the repulsion motion is contrasted by attracting forces modelled by linear springs between two
particles that share an edge.
Conventionally, the drawing is \emph{nice} when the system is at
the equilibrium. Nodes are drawn at the position reached by the
corresponding particles and straight lines are drawn between
nodes connected by an edge in the original graph.

Over the years also the aesthetic criteria for
graph drawing have evolved. Currently,
several criteria are commonly accepted as characterising~\cite{drawing_aesthetics_purchase,perceptual_organization_in_user_generated_grpah_layouts}
a \emph{nice} graph drawings, such as the
angular resolution, the distribution of vertices in the plane, or the number of edge crossings.

This paper focuses on the last property. Indeed, we aim at finding a drawing which minimizes the number of crossings when the edges are drawn as straight lines. We call this problem the \emph{rectilinear crossing minimization} problem (RCM problem for short).

RCM is a known computationally difficult problem. Indeed, solving
RCM for a generic graph is complete for the existential theory
of the reals and hence its
complexity (in the classical setting) is somewhere between \NP\ and \PSPACE~\cite{shaefer-complexity-of-rectilinear-crossing-problem}.

As a consequence, it is a natural research direction to look for
heuristics proposing trade-offs between exact solutions and
computational time.

%ICI IL FAUT METTRE LES TRUCS DES ALLEMANDS.... D'AUTRES ?
%Some strategies have been developped to move iteratively vertices to other positions of $\R^2$. Radermacher et al.~\cite{geometric_heuristics_for_rectilinear_crossing_minimization} proposed a way, according to a vertex of the graph, to divide the plane into regions in which the vertex have the same number of crossings wherever ont the region, and hence find the optimal one.\hl{ATTENTION: ce dernier paragraphe est mal écrit... à revoir!}
Some strategies have been developped to iteratively move vertices to other positions in $\R^2$. Radermacher et al.~\cite{geometric_heuristics_for_rectilinear_crossing_minimization,vertex_movement_radermacher_rutter} proposed a way, given a rectilinear drawing, to find for any vertex $v$ its optimal position, keeping all the other vertices fixed. To the best of our knowledge, this algorithm provides the best trade-off between precision and time.

%ICI IL FAUT DIRE CE QUE NOUS APPORTONS DE NOUVEAU + INTUITION ET INSPIRATION DE L'ALGO
Similarly to the Radermacher et al. approach,
the main strategy of our algorithm consists in improving an already existing drawing of a graph, step by step, moving a vertex to another position, potentially decreasing the number of crossings.
However, contrary to the computation of the optimal place which is rather costly, our goal is to find satisfying positions with cheaper mechanisms. To do so, the idea is to cast rays from a vertex.
These rays can either reflect on the edges, or go through them, according to a wisely chosen score function.
After several reflections or traversals, the ray ends up in a place which defines a positions where the vertex can be evaluated to move in or not.
The insurance of obtaining a satisfactory position comes from a theorem of the dynamical systems which states that, in a rectangular billiard, a trajectory with an irrational angle is dense in the space. For our case, it gives the intuition that our rays can find the optimal position since they can go through any open subset of the space.

%ICI IL FAUT DIRE QUELS SONT NOS RESULTATS
This algorithm has shown interesting results compared to the best rectilinear drawing algorithms of \emph{OGDF}, the reference library for this domain. It can also easily be tuned with various parameters to privilege either the quality of the results or the computation time.

The paper is structured as follows. In the next section,
all basic definitions and concepts are introduced.
Section~\ref{sec:heu-explanation} introduces and explains the new heuristics, while Section~\ref{sec:param} discusses the heuristic parametrization.
Convergence and complexity are discussed
in Section~\ref{sec:properties}. Finally, experimental results are shown in Section~\ref{sec:perfs} asserting the relevance of the parameters of the algorithm and comparing its performances with its main (available) competitor.

% !TEX root = ./main.tex

%In topological graph theory, an embedding (also spelled imbedding) of a graph
%G on a surface \Sigma  is a representation of  G
%on  \Sigma  in which points of \Sigma  are associated with vertices and simple arcs (homeomorphic images of
%[
%0
%,
%1
%]
%[0,1]) are associated with edges in such a way that:
%
%the endpoints of the arc associated with an edge
%e
%e are the points associated with the end vertices of
%e
%,
%e,
%no arcs include points associated with other vertices,
%two arcs never intersect at a point which is interior to either of the arcs.
%Here a surface is a compact, connected
%2
%2-manifold.

\section{Definitions}\label{sec:defs}

A graph $\GG$ is a structure $\structure{V,E}$
where $V$ is the (finite) set of \emph{vertices}
and $E\subseteq V\times V$ is the set of \emph{edges}. For any $v\in V$, $E_v$ is
the maximal subset of $E$ such that if $(a,b)\in E_v$,
then either $a=v$ or $b=v$.

An \emph{embedding} $\Pi$ of $\GG$ in a surface $\Sigma$
%(a compact connected $2$-manifold)
is a representation of $\GG$ in which
$V$ are points in $\Sigma$ and edges are simple
curves over $\Sigma$ (homeomorphic to $[0,1]$).
Moreover, the representation must be such that
(1) endpoints of a curve associated with an edge
must coincide with the endpoints of the edge; (2) no curve representing an edge contains more than two vertices;
(3) no two curves (representing edges) intersect at a common interior point.
A \emph{straight-line drawing} $\Gamma$
of a graph $\GG$ is an embedding of $\GG$ into $\R^2$
in which condition (3) is relaxed and edges are not
associated with generic curves but with straight-line segments.
Hence, to describe a straight-line drawing of $\GG$ one just needs to provide a bijective map from $V$ to $\R^2$.

A graph is \emph{planar} if it admits an embedding in $\R^2$.
The Fáry's theorem~\cite{planar_Fary} states that for any planar graph
there exists a straight-line drawing without crossing
edges. On the other hand, if $\GG$ is not planar,
then any straight-line drawing will have some crossing
edges. %It is a challenging problem to design
%algorithms which find straight-line drawings with
%a minimal number of crossings.

Given two edges $(a,b)$ and $(c,d)$ in $E$ and
a drawing $\Gamma$ of a graph \GG,
% we say that they \emph{cross} each other if the segments $\overline{\Gamma(a)\Gamma(b)}$ and $\overline{\Gamma(c)\Gamma(d)}$ intersect in a point which is not an endpoint of one of the two edges.
we denote the fact that they cross each other by
$(a,b)\crosses(c,d)$ without explicit reference to the dependency on $\Gamma$ when the embedding is clear from the context.
Therefore the \emph{crossing number} $cr(e)$ of an edge $e\in E$ is given by
$\card{\set{e' \in E\backslash\set{e}\,\mid\,e \times e'}}$. Intuitively, the crossing
number of a vertex is $cr(v)=\sum_{e\in E_v} cr(e)$ and that of a drawing $\Gamma$ is $cr(\Gamma)=\frac{1}{2}\cdot\sum_{e\in E} cr(e)$.
% Therefore the \emph{crossing number} $cr(v)$ of a node $v\in V$ is given by
% \card{\set{(a,b)\in E\,|\,\exists u\in V, (v,u)\crosses(a,b)}}. Finally, the crossing
% number of a drawing $\Gamma$ is $cr(\Gamma)=\frac{1}{2}\cdot\sum_{v\in V} cr(v)$.

We also define the function $\energy[v]$, called the \emph{energy} of $v$, as the sum of the squared norms of the Hooke's law forces applied to the endpoints of $(a,b) \in E_v$. Recall that this law models spring forces and can be seen as the delta between the actual length of the edge and a desired theoretical one. Similarly to the crossings function, $\energy[e]$ is the energy of one edge $e$ and $\energy[\Gamma]$ is the sum of the energies of all the edges in $\Gamma$.

The \emph{faces} of an embedding of a graph $\GG$ on a surface are the regions that remain when the points representing the vertices and edges of $\GG$ are removed from the surface. Remark that this kind of definition makes sense only for graphs for which we have found an embedding. Hence, in the sequel, we prefer the notion of \emph{facet} which, in a sense, describes the `real' visual faces. Following~\cite{planarization_book}, given a straight-line drawing $\Gamma$ of a graph $\GG$,  a \emph{planarization} $\Gamma'$ of $\Gamma$ can be obtained by replacing consecutively each pair of crossing edges by four new edges attached to an also new false vertex at the position of the old intersection point. The faces of this $\Gamma'$ do not overlap and are called the facets of $\Gamma$.

The \emph{bounding box} $\boundingBox$ of a graph $\GG$ is the minimum (\wrt surface) rectangle, aligned on the $x$ and $y$ axes, which contains all of the vertices of $\GG$ (considering that vertices have a null radius). We define also $\boundingBox[\varepsilon]$ as the \emph{expanded bounding box} of $\GG$ with a margin $\varepsilon$. We can visualize it as a rectangle with the same centroid and the same orientation as $\boundingBox$ but with a width (resp., a height) of length $w+2\varepsilon$ (resp., $h+2\varepsilon$) where $w$ (resp., $h$) is the width (resp., the height) of $\boundingBox$.

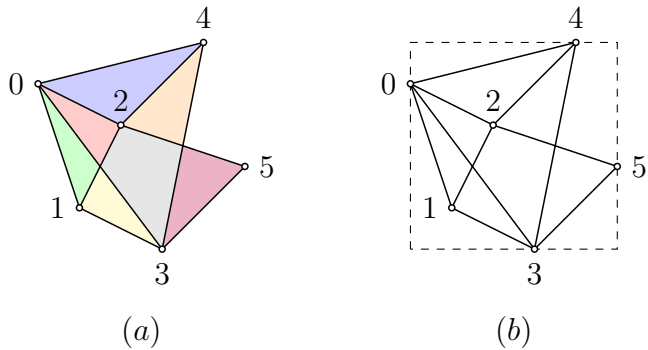
\begin{figure}
\centering
\begin{tikzpicture}[scale=.55]
\begin{scope}[xshift=-4.5cm]
\fill[blue!20] (0,4) -- (2,3) -- (4,5) -- (0,4);
\fill[red!20] (0,4) -- (2,3) -- (1.5,2) -- (0,4);
\fill[green!20] (0,4) -- (1,1) -- (1.5,2) -- (0,4);
\fill[yellow!20] (3,0) -- (1,1) -- (1.5,2) -- (3,0);
\fill[orange!20] (3.5,2.5) -- (2,3) -- (4,5) -- (3.5,2.5);
\fill[gray!20] (3.5,2.5) -- (3,0) -- (1.5,2) -- (2,3) -- (3.5,2.5);
\fill[purple!30] (3,0) -- (5,2) -- (3.5,2.5) -- (3,0);
\node[gnode] (0) at (0,4){};
\node[gnode] (1) at (1,1){};
\node[gnode] (2) at (2,3){};
\node[gnode] (3) at (3,0){};
\node[gnode] (4) at (4,5){};
\node[gnode] (5) at (5,2){};
\node[draw=none] (6) at (1.5,2){};
\node[draw=none] (7) at (3.5,2.5){};
\draw[gedge] (4) -- (0) -- (2) -- (4) -- (3) -- (5) -- (2) -- (1) -- (0) -- (3) -- (1);
\foreach \i/\p in {0/left,1/left,2/above,3/below,4/above,5/right}{
    \node[\p=0 of \i](){\i};
}
\node at (2.5,-2){$(a)$};
\end{scope}
\begin{scope}[xshift=4.5cm]
    \draw[style={dashed,color=black}] (0,0) -- (0,5) -- (5,5) -- (5,0) -- cycle;
    \node[gnode] (0) at (0,4){};
    \node[gnode] (1) at (1,1){};
    \node[gnode] (2) at (2,3){};
    \node[gnode] (3) at (3,0){};
    \node[gnode] (4) at (4,5){};
    \node[gnode] (5) at (5,2){};
    \draw[gedge] (4) -- (0) -- (2) -- (4) -- (3) -- (5) -- (2) -- (1) -- (0) -- (3) -- (1);
\foreach \i/\p in {0/left,1/left,2/above,3/below,4/above,5/right}{
    \node[\p=0 of \i](){\i};
}
\node at (2.5,-2){$(b)$};
\end{scope}
\end{tikzpicture}
\caption{The facets (a) and the bounding box (b) of a straight-line drawing $\Gamma$ of \GG.}
\label{fig:the-facets-of-G}
\end{figure}

Finally, we call \emph{ray} a polyline, expressed as a sequence of points $p_0,p_1,\ldots,p_r$ with $r>0$ and a half-line whose initial point is $p_r$ and its direction vector $\vv{d}$. In the sequel, we will say that \emph{a ray intersects an edge} of a graph $\GG$ if the half-line of the ray intersects it, but not the polyline. The length of the sequence will be called the size of the ray.

\begin{figure}[ht]
\centering
\begin{tikzpicture}[scale=.55]
    \node[gnode] (0) at (1,3){};
    \node[gnode] (1) at (5,5){};
    \node[gnode] (2) at (2,6){};
    \node[gnode] (3) at (3,0){};
    \node[gnode] (4) at (5,1){};
    \node[gnode] (5) at (1,1){};
    \node[above=0 of 0] (){$p_0$};
    \node[right=0 of 1](){$p_1$};
    \node[above=0 of 2](){$p_2$};
    \node[below=0 of 3](){$p_3$};
    \node[right=0 of 4](){$p_4$};
    \node[left=0 of 5](){$p_5$};
    \node[draw=none](v1) at (1.5,1.25){};
    \node[above=-.1 of v1](){$\vv{d}$};
    \draw[-stealth](5) -- (2,1.5);
    \draw[gedge] (0) -- (1) -- (2) -- (3) -- (4) -- (5);
    \draw[gedge] (5) -- (4,2.5);
    \draw[gedge, style={dashed}] (5,3) -- (4,2.5);
\end{tikzpicture}
\caption{An example of a ray.}
\label{fig:ray}
\end{figure}
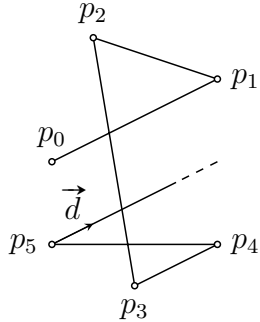

% !TEX root = ./main.tex

%
\section{The new heuristic}\label{sec:heu-explanation}
We propose a new heuristic for rectilinear graph
drawing called ``Ray-based Rectilinear Graph Drawing"
(\RRGD).
% Its pseudo-code is given in Algorithm~\ref{alg:rrgd}. This section details it step by step.
\medskip

\RRGD takes in input a finite graph $\GG = \structure{V,E}$ and calls \texttt{Init(G)} %(line~\ref{alg:rrgd:init})
to get
an initial drawing $\Gamma_0$. The possible intial drawings are completely independant of the algorithm, meaning that the result is not sensible to $\Gamma_0$, and will be discussed in Section~\ref{sec:drawingInit}. A drawing $\Gamma$
is represented by a list of pairs $(\texttt{v},\texttt{p}_\texttt{v})$ where $\texttt{v}$ is the name of the
node and $\texttt{p}_\texttt{v}=(\texttt{v}_x,\texttt{v}_y)$ are its coordinates.
%The list
%is sorted according to $cr(\texttt{v})$ first and then
%to $F_\texttt{v}$.

%\begin{algorithm}
%\caption{$updateDrawing(\Gamma)$} \label{alg:global}
%\begin{algorithmic}
%\Require $\GG = \structure{V,E}$
%\State $L \gets nodeSort(V)$ \Comment{Using $(cr(v),\energy[v])$ as key for comparison}
%\State $update \gets True$
%\While {$update$}
%\State $update \gets False$
%\For{$v \in L$}
%    \State $p \gets pos(v)$
%    \State $p\prime \gets findBetterPosition(v)$
%    \If{$p \neq p\prime$}
%        \State $pos(v) \gets p\prime$
%        \State $update \gets True$
%        \State $\text{updating }L$
%        \State \textbf{break}
%    \EndIf
%\EndFor
%\EndWhile
%\end{algorithmic}
%\end{algorithm}

The \texttt{Init} function also  sets up a bounding box $\boundingBox[\varepsilon]$ with four dummy vertices, along with four dummy edges. The latter are fixed and will not be moved during the whole execution of the algorithm. The box $\boundingBox[\varepsilon]$ will help to keep the other vertices in a reasonable frame.
Also, the box is set with a margin $\varepsilon>0$ to give to the  real vertices more degrees of freedom to move to the other side of $\Gamma$ by going around the whole graph using the gap between vertices and the dummy edges.
%Indeed, the algorithm does not necessarily need a \emph{good} initial guess for $\Gamma$ to start with.
% As we will see in the experiments later, the results of the algorithm are not sensible to $\Gamma_0$ but it is clear that a decent initial guess can make it converge faster.

After the initialization, the algorithms enters its main
loop which keeps running as long as the
\texttt{Move} function finds a better position for at least one vertex; otherwise it stops and a $\Gamma$ is returned.
We will call $\Gamma_n$ the drawing of $\GG$ produced after $n$ iterations of the main loop.

\subsection{Algorithm explanation}
\paragraph{Sorting vertices} \label{subsec:vertex_sorting}

At each run of the main loop, the vertices of $\GG$ are
sorted in descending order according to their position in $\Gamma$ to treat problematic vertices first. To do so, we define the order
$\leq_\Gamma$ as follows: for any pair of
vertices $u,v\in V$, $u\leq_\Gamma v$ if
$cr(u)<cr(v)$ or, in case $cr(u)=cr(v)$, $\energy[u] \leq \energy[v]$.

% \hl{
% Recall that the Hooke's law is given by $\vv{F}=-k\vv{\Delta l}$, where $\vv{F}$ is the force of traction or compression, $k$ is a constant (which depends on the spring) and $\vv{\Delta l}$ is the variation of the spring length with respect to a default length $\vv{l_0}$. From the physics point of view, $\vv{l_0}$ is the spring length at rest.
% }

The consideration about the vertices energy allows to favor edges of homogeneous sizes.
% This is another commonly used criterion for a graph drawing to be considered as \emph{pleasingly} looking~\cite{perceptual_organization_in_user_generated_grpah_layouts} \fd{ref à discuter}.
Although the repartition of edge lengths is not the most valued criterion to qualify a drawing as \emph{pleasingly} looking~\cite{perceptual_organization_in_user_generated_grpah_layouts}, it is nevertheless often considered when talking about graph drawing.
Moreover, this metric is implicitly
used in all force-based models since it represents
the spring length.

\begin{algorithm}[ht!]
\caption{\texttt{RRGD}}\label{alg:rrgd}
\begin{algorithmic}[1]
\Require $\GG = \structure{V,E}$
\State $\Gamma=$ \texttt{Init(G)} \label{alg:rrgd:init}
\State $update \gets true$
\While {$update$} \label{alg:rrgd:first_loop_start}
\State $update \gets false$
\State $\Gamma \gets \texttt{Sort}(\Gamma)$
\For{$(\texttt{v},(\texttt{v}_x,\texttt{v}_y)) \in\Gamma$} \label{alg:rrgd:second_loop_start}
    \State $(\texttt{v}\prime_x,\texttt{v}\prime_y) \gets \texttt{Move}(\texttt{v},\Gamma)$
    \If{$(\texttt{v}_x,\texttt{v}_y) \neq (\texttt{v}\prime_x,\texttt{v}\prime_y)$}
        \State $\Gamma(\texttt{v}) \gets (\texttt{v}\prime_x,\texttt{v}\prime_y)$
        \State $update \gets true$
        \State \textbf{break}
    \EndIf
\EndFor \label{alg:rrgd:second_loop_end}
\EndWhile \label{alg:rrgd:first_loop_end}
\State\Return $\Gamma$
\end{algorithmic}
\end{algorithm}

%Thanks to these criteria, the first focus will be on the vertices, surely which crosses a great number of edges, but also vertices which have disproportionated edges, either very long or very short.

\paragraph{The inner loop and the \texttt{Move} function}

The inner loop spans through the pairs
$(\texttt{v},\texttt{p}_\texttt{v})$ of $\Gamma$ calling \texttt{Move} to check if $\Gamma$ can be
improved.
% The pseudo-code for \texttt{Move} is given in Algorithm~\ref{alg:move}.
Given a vertex $v$ and a drawing $\Gamma$,
\texttt{Move} builds a list $L$ of candidate positions
by calling \texttt{CastRay} $R$ times with different angles and returns the minimum (according to $\leq_\Gamma$) of $L$. %Remark that at the end of the for loop (line~\ref{alg:move:endfor}), $L$ has length $R+1$ (where $R$ is a parameter defining the number of rays cast and the $+1$ accounts for the current position which is compared with the new potential ones).

%Moreover, to get closer to our main theorem, we initialize each ray with an irrational base angle (or a floating approximation of an irrational number) $\theta_0$. This allow us to minimize also the chances to fall in special cases, such as having rays passing through vertices, or hitting an edge perpendicular and be trapped in two parrallel edges. Both of these cases can occur more frequently if the base angle of the rays are a fraction of $\pi$ and if the vertices are initialized on a grid or in circle for instance.

% Each angle is initialized with the same constant $\theta_0$ which represents an irrational angle. \todo{En commentaire dans le pseudo code}

% In this way, each ray has a non-null probability
% to pass close to any point of the plane
% \hl{ici il faudrait renvoyer sur un résultat dans lequel
% on montre ceci ou on explique au moins.}

\begin{algorithm}[ht!]
\caption{$\texttt{Move}(\texttt{v},\Gamma)$} \label{alg:move}
\begin{algorithmic}[1]
\State $L \gets [\Gamma(v)]$
\For{$i_r \in [1..R]$}
    \State $\theta \gets \theta_0 + i_r \times \frac{2\pi}{R}$ \label{alg:move:theta} \Comment{$\theta_0$ is chosen irrational to avoid some corner cases.}
    \State $(p_x,p_y) \gets \texttt{CastRay}(\texttt{v},\theta, \Gamma)$
    \State $L.append((p_x,p_y))$
\EndFor
\State $\textbf{return } \texttt{min}_{\leq_\Gamma}(L,\Gamma)$ \label{alg:move:endfor} %\Comment{Using $(cr(v),\energy[v])$ as key for comparison}
\end{algorithmic}
\end{algorithm}

\paragraph{Ray casting} \label{par:ray_casting}

\begin{figure}
\centering
\begin{tikzpicture}[scale=.26]
 \draw[ray] (0,0) -- (12.5,12.5);
 \draw[ray] (0,0) -- (-8.5, 2.28);
 \draw[ray] (0,0) -- (2, -7.5);
 \node[gnode,style={draw=jblue}] (0) at (0,0){};
 \node[gnode] (1) at (0,5){};
 \node[gnode] (2) at (5,0){};
 \node[gnode] (3) at (0,10){};
 \node[gnode] (4) at (-5,5){};
 \node[gnode] (5) at (-5,0){};
 \node[gnode] (6) at (0,-5){};
 \node[gnode] (7) at (5,-5){};
 \node[gnode] (8) at (10,0){};
 \node[gnode] (9) at (10,5){};
 \node[gnode] (10) at (5,10){};
 \draw[gedge] (1) -- (3) -- (1) -- (4) -- (1) -- (2) -- (7) -- (2) -- (8);
 \draw[gedge] (3) -- (4) -- (5) -- (6) -- (7) -- (8) -- (9) -- (10) -- (3);
 \draw[gedge,style={color=jblue}] (1) -- (0) -- (2) -- (0) -- (9) -- (0) -- (10) -- (0) -- (6);
 \node[gnode] (c0) at (-8.5,12.5){};
 \node[gnode] (c1) at (-8.5,-7.5){};
 \node[gnode] (c2) at (13.5,-7.5){};
 \node[gnode] (c3) at (13.5,12.5){};
 \draw[gedge,style={dashed}] (c0) -- (c1) -- (c2) -- (c3) -- (c0);
 \node[style={text=red}] at (2.5,2.5){$\bullet$};
 \node[style={text=brickRed}] at (7.5,7.5){$\bullet$};
 \node[style={text=brickRed}] at (12.5,12.5){$\bullet$};
 \node[style={text=brickRed}] at (-8.5, 2.28){$\bullet$};
 \node[style={text=red}] at (-5, 1.34){$\bullet$};
 \node[style={text=brickRed}] at (2, -7.5){$\bullet$};
 \node[style={text=red}] at (1.34,-5){$\bullet$};

 \node[style={text=jblue}] at (-1,-1){$v$};
 \node[] at (3.9,2.7){$p_{0}$};
 \node[] at (8.9,7.5){$p_{1}$};
 \node[] at (12.5,11.4){$p_{2}$};
 \node[style={text=red}] at (9.5,10.5){$r_0$};
 \node[style={text=red}] at (-6.75,1.2){$r_1$};
 \node[style={text=red}] at (1,-6.25){$r_2$};
 \node[] at (-7.2,11.5){$\boundingBox[\varepsilon]$};
\end{tikzpicture}
\label{fig:rays}
\caption{Rays $r_0, r_1$ and $r_2$ casted from vertex $v$ and their respective intersection points.}
\end{figure}
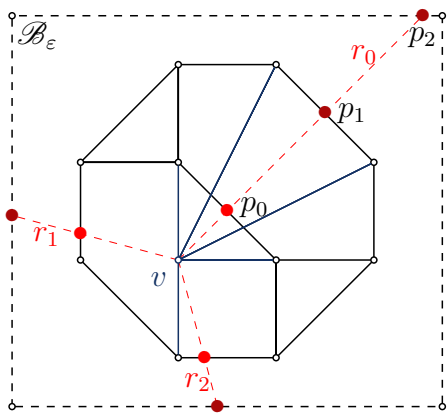

For one initial position $\texttt{p}_\texttt{v}$ and one angle $\theta$, the algorithm consider a half-line matching these parameters. It then computes the intersections points with the real edges of $\GG$ and also with the four dummy ones representing $\boundingBox[\varepsilon]$. These intersection points are then sorted according to their distance to $\texttt{p}_\texttt{v}$ (with the closer ones first) and be processed in this order for the next step. Figure~\ref{fig:rays} shows three rays being cast from one node and intersecting the edges of the graph and its bounding box.

%The main idea to move a vertex $v$ to other places is to cast several rays from the initial position of $v$ and see where they end up. Thus, at the start of each iteration, we begin by casting a number $R$ rays evenly spread angle-wise, in other words, we construct each ray with a sequence of only one point $p_0$ which will be equal to the position of $v$ and with for each a different angle of $\vv{d}$.

%These rays, more precisely the half-lines, will intersect some edges of $\GG$ and one added edge of $\boundingBox_m$. Let denote by $p_i$ the intersection points of the ray with the edges encountered. There is always at least one $p_i$ since $v$ is inside $\boundingBox_m$ and thus is sure to cross exactly one of the four added edges (we consider the probability for a ray to pass exactly through a corner of $\boundingBox_m$ as 0). Then, we choose the closest $p_i$ from $v$, with euclidean distance, and consider it for the potential point of reflection.

% To move a vertex $v$ to a better location, we start by casting $R$ rays from $v$. For each ray, we compute all the intersections with the edges of the graph $\GG$ plus four arbitrary false edges which delimits the bounding box of $\GG$. We keep the closest one and go to the second step. For example, if $R=3$ and the node $v$ is considerd. For $r_0$, all the intersection points with $\GG$ or the 4 added edges (dotted lines), are $p_{00}$, $p_{01}$ and $p_{02}$. The only point considered for the next step for this ray is $p_{00}$, which is the first one met starting from $v$.

%But du lemme:

%\subsection{Edge Opacity}

\paragraph{Crossing or reflecting}
When a ray hits an edge $e$, it can either
pass through or reflect on it according to the \emph{opacity} of $e$. This quantity measures
the average decrease (or increase) of the crossing number of a node whenever it is moved before or beyond an intersection point. Its value depends on three parameters: the vertex $v$ which we try to move, the edge $e$ that the current ray has crossed, and an evaluation position $p$ of $v$. We define $p \prime$ and $p \prime \prime$ as the points on the ray $r$ at a distance $\varepsilon$ from the intersection point of $r$ and $e$, with $p \prime$ being the closest to the last added point of the sequence of points of $r$ (see Figure~\ref{fig:weight-attribution}).

To compute the opacity, we need first to assign to each edge $e_i \in E_v$ a weight $w_{e_i}$ %\hl{given the configuration where $v$ is temporarily placed at $p \prime$.} \hl{la phrase précédente est incomprensible...} The weights are defined
as follows:

% When a ray, cast from a node $v$ is about to cross an edge $e$, a value is attributed to $e$ to determine if the ray goes through the edge or if it reflects. We call this value \emph{edge opacity} and denote it by $\texttt{Opacity}(v,e)$. To compute it, we start by assigning to each edge $e_i \in E(v)$ a weight $w_{e_i}$ as follows:

\begin{equation*}
w_{e_i} =
\begin{cases}
      \phantom{-}0, & \text{if $e_i$ shares a vertex with $e$,}\\
     -1,            & \text{if $e_i$ does not share a vertex with $e$ and $e_i$ crosses $e$,}\\
      \phantom{-}1, & \text{if $e_i$ does not share a vertex with $e$ and does not cross $e$.}
\end{cases}
\end{equation*}

\begin{center}
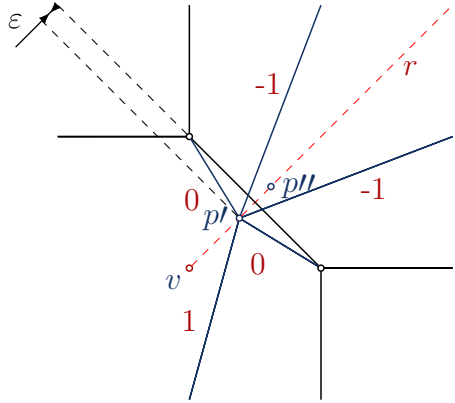

\begin{tikzpicture}[scale=.35]
    \node[gnode] (1) at (0,5){};
    \node[gnode] (2) at (5,0){};
    \node[gnode,style={draw=jblue}] (0) at (1.9,1.9){};
    \node[gnode,style={draw=brickRed}] (3) at (0,0){};

    \draw[ray] (3) -- (0) -- (10,10);
    \draw[gedge] (0,10) -- (1) -- (-5,5) -- (1) -- (2) -- (10,0) -- (2) -- (5,-5);
    \draw[gedge,style={color=jblue}] (1) -- (0) -- (2) -- (0) -- (0,-5) -- (0) -- (10,5) -- (0) -- (5,10);

    \node[label] at (2.5+0.1,0+0.18){0};
    \node[label] at (0+0.1,2.5+0.1){0};
    \node[label] at (7,3){-1};
    \node[label] at (3,7){-1};
    \node[label] at (0,-2){1};

    \node[style={text=jblue,fill=none},below left=-3pt of 3](){$v$};

    \node[gnode,style={draw=jblue}] (4) at (3.1,3.1){};
    \node[style={text=jblue,fill=none},right=-1pt of 4](){$p \prime \prime$};
    \node[style={text=jblue,fill=none},left=-1pt of 0](){$p \prime$};

    \node[label]() at (8.4,7.6){$r$};

    \draw[style={dashed,color=black}] (1) -- (-5,10);
    \draw[style={dashed,color=black}] (0) -- (-5.6,9.4);

    \draw[latex reversed-latex reversed,style={line width=0.18mm}] (-5,10) -- (-5.6,9.4);
    \draw[style={line width=0.18mm}] (-5.6,9.4) -- (-6.6,8.4);
    \node[] at (-6.6,9.4){$\varepsilon$};
\end{tikzpicture}
\captionof{figure}{An example of weight attribution for the computation of opacity. When trying to move
$v$ from $p\prime$ to $p\prime\prime$, all the edges will have opposite weights in $p\prime\prime$ (this actually holds whatever the edge configuration except for the positive edges, which can remain positive in $p\prime\prime$ if a neighbor of $v$ is co-linear with the endpoints of $e$).}
\label{fig:weight-attribution}
\end{center}

%These weights help us to determine the benefit to pass from $p \prime$ to $p \prime \prime$, i.e., cross the edge.

Then, the opacity is the average of the weights, ignoring null values:
% After we computed the weights, the \emph{opacity} of $e$ according to $v$ is defined as

\begin{equation*}
\text{\texttt{Opacity}}(v,e,p) =
\begin{cases}
      \ffrac{\sum\limits_{e_i \in E_v} w_{e_i}}{\card{E^*_v}}, & \text{if }\card{E^*_v}>0\\
      \phantom{-...... .  ....}1, & \text{otherwise}
\end{cases}
\end{equation*}
where $E^*_v$ is the subset of $E_v$ for which
each edge $e_i$ has $w_{e_i}\ne0$.

Finally, remark that in the special case in which $e$ is one of the four sides of $\boundingBox[\varepsilon]$, the opacity is not computed, since the ray reflects on it by default.
\smallskip

%\subsection{Iteration}

\noindent
At the end of the opacity calculation, two ``actions" can be taken:
\medskip

\begin{minipage}{.9\textwidth}
\paragraph{Reflection}
The point $p$ is added to the sequence of the ray and the direction vector $\vv{d}$ is updated as follows. Let $\theta_d$ and $\theta_e$ be respectively the angles of $\vv{d}$ and $e$, then the we construct a new unit vector $\vv{d}$ with an angle $\theta=2*\theta_e - \theta_d$. Note that since $\theta$ is taken modulo $2\pi$, the direction of the edge $e$ does not matter.

%Let suppose that we choose one direction of $e$. This gives us one $\theta_e$, which can be used to compute our new $\theta$. If we would have considered the other direction, and thus consider another angle $\theta_e^\prime$. $\theta_e^\prime$ would equals $\theta_e + \pi$, and we would have $\theta^\prime = 2*\theta_e + 2\pi - \theta_d$, which equals $\theta$ modulo $2\pi$.

\paragraph{Crossing.}

The point $p$ is also added to the sequence of the ray (even though it can be collinear to the two previous points in that sequence). However, the unit vector $\vv{d}$ remains unchanged. In this case, there is no need to recompute the intersections points of the ray with the edges, we can only take the next one in the sorted list and recompute its opacity with its associate edge. Note that $p$ is also added to the sequence of the ray even though it will be co-linear with its predecessor and its successor in the sequence.

%We take the next intersection point computed in the "\nameref{par:ray_casting}" step and recompute the opacity with another $e$.

%We also introduce an escape strategy to speed up computation time.  If the ray is reflected $n_{esc}$ times in a row we stop the ray here and compute the next one. Most of the time, the consecutive reflections means that $v$ is stuck in a reasonable facet. This facet is a local minima, it surely exists a way to escape this facet to go to a better one but the chances are to small to pay the cost of the following applications of the routine to do it $n_r$ times.
\end{minipage}

\noindent
For now, we only consider that the ray reflects if there are more edge with a $1$ than with a $-1$.
We keep doing this routine $n_r$ times to let it visit through a non-negligible portion of $\Gamma$.

\paragraph{Pseudocode}

% The \texttt{move} function calls \texttt{castRay}
% for determining a new candidate position for a node $v\in V$.
The ray is casted at position $\Gamma(v)$
according to an angle $\theta$ and built point by
point. At each run of the main loop (line~\ref{alg:ray:main_loop}) a
new point it is added. Let $p$ be the current point
of the ray that have just been built. The next one is chosen among the intersection points $p_i$ which
are given by the half-line exiting from $p$ with angle $\theta$ and one of the edges (via calls to the function \texttt{intersectionHalflineSegment} at line~\ref{alg:ray:intersect}). Lines \ref{alg:ray:heap_loop}-\ref{alg:ray:heap_loop_end}
arrange such candidates on a heap (which uses the Euclidean distance between $p$ and $p_i$ as key for comparison between points). Remark that
after the execution of the lines \ref{alg:ray:heap_loop}-\ref{alg:ray:heap_loop_end}, the heap
$H$ is always non empty because of the bounding
box $\boundingBox[\varepsilon]$.
Hence the top of the heap can be safely popped
(line~\ref{alg:ray:pop}) into $p'$ and $p'$ is added to the ray.
At this point (lines \ref{alg:ray:refl_if}-\ref{alg:ray:refl_if_end}) the
algorithm decides if the ray is going to be reflected
at $p'$ or it pass through the edge to which $p'$
belongs to. If the heap $H$ is empty then it means
that $p'$ is a point on one of the edges of
$\boundingBox[\varepsilon]$ and hence the ray must
reflect. This is also the case if the opacity of
the edge (computed by the \texttt{Opacity} function
at line \ref{alg:ray:refl_if}) is less or equal than zero. Finally,
the procedure returns the mid-point of the last
segment of the ray. Remark the point returned is
chosen in this way to place the node relatively
close to the centers of the facets. This also allows
to ensure a higher angle resolution in most cases at a minimum cost.

\begin{algorithm}[ht!]
\caption{$\texttt{CastRay}(\texttt{v},\theta, \Gamma)$\hfill (Deterministic mode)} \label{alg:castRay}
\begin{algorithmic}[1]
\State $p \gets \Gamma(v)$
\State $H \gets heap()$
\State $Ray \gets [p]$
% \State $n_{refl} \gets 0$
\For{$i \gets 1$ to $n_r$} \Comment{$n_r$ is a constant fixed in advance} \label{alg:ray:main_loop}
    \If{$H.isEmpty()$}
        \For{$e \in E$} \label{alg:ray:heap_loop}
            \State $p_i \gets intersectionHalflineSegment(p,\theta,e)$ \label{alg:ray:intersect}
            \If{$p_i \neq null$}
                \State $H.push(p_i)$ \Comment{Using $dist(p,p_i)$ as key for comparison}
            \EndIf
        \EndFor \label{alg:ray:heap_loop_end}
    \EndIf
    \State $p\prime \gets H.pop()$ \Comment{$\varepsilon$-distance omitted here} \label{alg:ray:pop}
    \State $Ray.append(p\prime)$
    \State $e \gets associateEdge(p\prime)$
    \If{$H.isEmpty()$ or $\text{\texttt{Opacity}}(\texttt{v},e,p\prime) \leq 0$} \label{alg:ray:refl_if}
        \State $p \gets p\prime$
        \State $\theta_e \gets edgeAngle(e)$
        \State $\theta \gets 2 \times \theta_e - \theta$
        % \State $n_{refl} \gets n_{refl}+1$
        \State $H.clear()$
    % \Else
        % \State $n_{refl} \gets 0$
    \EndIf \label{alg:ray:refl_if_end}
    %\If{$n_{refl} == n_{esc}$}
        % \State \textbf{break}
    % \EndIf
\EndFor
\State $p_1 \gets Ray.pop()$
\State $p_2 \gets Ray.pop()$
\State \textbf{return} $\frac{p_1+p_2}{2}$ \Comment{positions $p_1$ and $p_2$ are 2D-vectors}
\end{algorithmic}
\end{algorithm}

\begin{figure}
\centering
\begin{tikzpicture}
\begin{scope}[xshift=0cm]
\draw[ray](3.60,2.54) -- (4.00,2.96) -- (3.03,4.00) -- (2.13,3.03) -- (1.69,2.56) -- (1.56,2.42) -- (2.35,2.34) -- (2.99,2.27) -- (2.87,0.41);
\foreach \i/\x/\y in {0/3.77/1.34,1/3.60/2.54,2/3.03/0.40,3/1.98/1.29,4/2.52/2.82,5/1.53/2.51,6/1.01/3.60,7/0.23/1.85,8/1.38/0.55}{
\node[gnode](\i) at (\x,\y){};
}
\foreach \u/\v in {0/1,0/2,0/4,1/2,1/4,2/8,3/4,3/5,3/8,4/5,4/6,5/8,5/6,6/7,7/8}{
\draw[gedge] (\u) -- (\v);
}
\node[gnode] (c0) at (0,0){};
\node[gnode] (c1) at (0,4){};
\node[gnode] (c2) at (4,4){};
\node[gnode] (c3) at (4,0){};
\draw[gedge,style={dashed}] (c0) -- (c1) -- (c2) -- (c3) -- (c0);
\node[gnode,fill=black]() at (2.93,1.34){};
\node[fill=none,draw=none]() at (3.60,2.74){$v$};
\node[fill=none,draw=none]() at (2,-.5){$\Gamma_i$};
\end{scope}
\begin{scope}[xshift=5cm]
\foreach \i/\x/\y in {0/3.77/1.34,1/2.93/1.34,2/3.03/0.40,3/1.98/1.29,4/2.52/2.82,5/1.53/2.51,6/1.01/3.60,7/0.23/1.85,8/1.38/0.55}{
\node[gnode](\i) at (\x,\y){};
}
\foreach \u/\v in {0/1,0/2,0/4,1/2,1/4,2/8,3/4,3/5,3/8,4/5,4/6,5/8,5/6,6/7,7/8}{
\draw[gedge] (\u) -- (\v);
}
\node[gnode] (c0) at (0,0){};
\node[gnode] (c1) at (0,4){};
\node[gnode] (c2) at (4,4){};
\node[gnode] (c3) at (4,0){};
\draw[gedge,style={dashed}] (c0) -- (c1) -- (c2) -- (c3) -- (c0);
\node[fill=none,draw=none]() at (2.73,1.34){$v$};
\node[fill=none,draw=none]() at (2,-.5){$\Gamma_{i+1}$};
\end{scope}
\end{tikzpicture}
\caption{The node $v$ in $\Gamma_i$ is moved to another position in $\Gamma_{i+1}$ after a ray has been cast from it (having been reflected 4 times and having crossed an edge 3 times). The ending position of the ray is shown as the black dot in $\Gamma_i$.}
\end{figure}
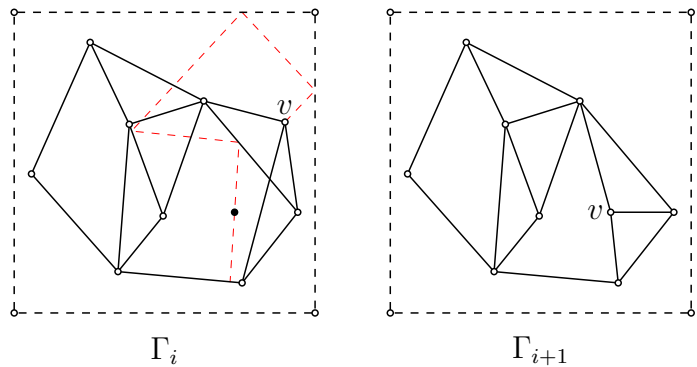

\paragraph{Position evaluation}

After repeating the \emph{``cross/reflect"} routine at most $n_r$ times, we end up with a sequence of length $n_r + 1$. We assign to each ray $r_i$ a final point $p_{r_i}$ as the midpoint of the last segment of the constructed polyline. Since $n_r >0$, it is always possible to find one. Then, we compare all the $p_{r_i}$ according to the same two criteria that we used in the ``\nameref{subsec:vertex_sorting}" section, except that this time, the lowest values are preferred.

We eventually move $v$ only if the best position among the $p_{r_i}$ is better than the initial position of $v$. If not, we apply the movement function to the next vertex in the list established at the begining of the pass.

%A \emph{ray} $r$ is a sequence $p_0, p_1, \ldots, p_{n_r}$, where $p_0$ is the intial position of $v$ and $p_i$, $i>0$ are the intersections or the reflection points of $r$ with the edges of $\GG$. For each ray, we define a final point $p$ as the middle of the last ray segment, \emph{i.e.}\ $p_{n_r-1}p_{n_r}$. Two criteria are used to find the best one among all the rays: first, the number of crossings of $v$ evaluated on the position $p$; second, the sum of the squared norms of the Hooke's law forces applied to the endpoints of the $e_i \in E(v)$ where $v$ is, again, evaluated on $p$.

% Finally, remark that lines $4$ and $23$-$28$ are
% there to perform a sanity check. Indeed, it might
% happen that the ray keeps reflecting between two
% parallel edges when its direction is perpendicular
% to them. For this reason we count the number of
% reflections ($n_{refl}$) and when a threshold is
% reached ($n_{esc}$) the extension of the ray is aborted.

Note that with the emphasis of the crossing number first for the comparison, a vertex with no crossing but with very long edges will always be preferred.
This is not an issue is the sense that there is a good chance that it is precisely these long edges that allow it to have no crossings (see Figure~\ref{fig:long_edge} in the appendix). This type of behavior seems to appear quite often.

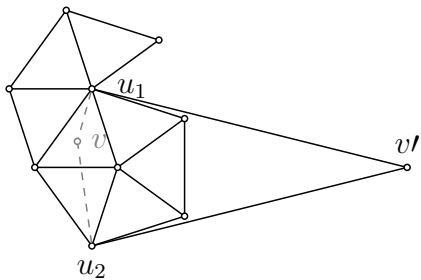
\begin{figure}
\centering
\begin{tikzpicture}[scale=.55]
\node[gnode](0) at (-0.62,-1.9){};
\node[gnode](1) at (-2,0){};
\node[gnode](2) at (0,0){};
\node[gnode](3) at (-0.62,1.9){};
\node[gnode](4) at (-2.62,1.9){};
\node[gnode](5) at (-1.24,3.8){};
\node[gnode](6) at (1.62,-1.18){};
\node[gnode](7) at (1.62,1.18){};
\node[gnode](8) at (7,0){};
\node[gnode](9) at (1,3.08){};

\foreach \u/\v in {0/1,0/2,0/6,0/8,1/4,1/3,1/2,2/3,2/6,2/7,3/4,3/5,3/7,3/8,4/5,6/7,3/9,5/9}{
    \draw[gedge] (\u) -- (\v);
}
\node[gnode,style={draw=gray}](10) at (-0.97,0.63){};

\draw[gedge,gray,dashed](0) -- (10);
\draw[gedge,gray,dashed](3) -- (10);
\node[right=0 of 10,gray](){$v$};

\node[above=0 of 8] (){$v\prime$};
\node[right=.15 of 3] (){$u_1$};
\node[below=0 of 0] (){$u_2$};
\end{tikzpicture}
\caption{An example of the need for long edges, where we want to find the minimum number of crossings by moving only the vertex $v$. Obviously, some better drawing of the graph above is possible but here, we place ourselves in the case where we can only move $v$ to another location and the rest of the graph stays put. Hence, the position $v\prime$ is near-optimal, in terms of crossing number and even for edge length.}
\label{fig:long_edge}
\end{figure}

%To move one node $v$ the algorithm runs in $O(kERn_r)$, where $E$ is the number of edges and $k$ is the degree of $v$. Moreover, at each iteration, it runs through the nodes until one can be moved to a better location. This leads to a total complexity of $O(E^2Rn_r)$ per iteration.

% !TEX root = ./main.tex

\section{Parametrization and refinement}\label{sec:param}

\subsection{Consideration of the opacity}

When a ray cast from a vertex $v$ is about to cross an edge $e$, then we compute the opacity of the edge given $v$, $e$ and an evaluation position $p$ placed on the ray at a distance $\varepsilon$ form the intersection point with $e$. This opacity will act like a score function to decide if the ray crosses $e$ or not. We define two ways to take into consideration the result of $\texttt{Opacity}(v,e,p)$.

\subsubsection{Deterministic reflections}

The first way to consider the opacity, as explained before, is to simply look at its sign. If it is negative, it basically means that more than half of the relevant edges, those which does not share a vertex with $e$, cross $e$.
% The deterministic mode relies only on the sign to let the ray cross $e$ or not.
In this case, with a negative opacity, we simply take the short-term best outcome and let the ray cross since it decreases the number of crossings. Thus, the ray reflects on $e$ if $\texttt{Opacity}(v,e,p)>0$. For the case where the opacity is null (\ie when there is no relevant edges or when there are as many edges that cross as those that do not), both actions can be defended. We chose to make the ray reflect in this case, to limit the number of positions for a vertex in an already satisfactory position and hence speed up the algorithm a bit.

% We will call \emph{conservative} the configuration of the algorithm when it makes reflect the ray when the opacity is null to try to keep $v$ in an acceptable position, and \emph{exploratory} when it tries to investigate different positions by making the ray cross $e$ when the opacity is null.

% \begin{lemma}
%     In conservative mode, if the hull of $\GG$ is convex, any vertex $v$ inside $\GG$ will stay inside and will preserve the hull.
% \end{lemma}

\subsubsection{Randomized reflections}

In this configuration, the opacity computed acts like a probability to cross or not. It will not directly determine the behavior of the ray but only bias it. We will compute for each intersection of a ray with an edge a random variable $\chi$ uniformly distributed on the interval $[-1-\varepsilon,1+\varepsilon]$. The ray will reflect if $\chi < \texttt{Opacity}(v,e,p)$, the higher the opacity is, the lower the chances to cross $e$ are. Note that thanks to $\varepsilon$ in the interval, both outcomes are always possible, even if all the edges have the same weight and $|\texttt{Opacity}(v,e,p)|=1$. %This allow us to explore \emph{less interesting} locations but to potentially have better results at the end. With this mechanism, we focus more on the long-term results than the short-term.
This will give the algorithm a chance to avoid local
optima and find global ones.

\subsection{Energy delta and Prohibition window}
We introduce now two parameters that each, in their own way, offer a balance between the computational time and the quality of the final drawing.

Firstly, the parameter $\deltaE$ is the amount of energy decrease necessary for a vertex $v$ to move to another location $p$ if $p$ does not improve the crossing number.
This threshold is mandatory to avoid some cases where the algorithm enters a loop,
where the succession of application of the node movement functions from a drawing $\Gamma$ leads eventually to itself.

Secondly, the parameter $\prohibitionWindow$ defines, for each vertex $v$ that has been moved, the number of iterations required before it can be moved again.
The purpose of this parameter is mainly to speed up the algorithm.
If the algorithm only moves the worst rated vertex (according to $\leq_\Gamma$) continuously,
we can quickly fall into a case where even after moving it, it stays the worst and the algorithm do not stop moving it bit by bit, thus neglecting all the other vertices.
For instance, if the vertex in question is trapped in a good facet (\ie it reflects on every edge which may not decrease the crossing number) but the movement function diminish only its energy.
Theoretically, a $\deltaE$ large enough could handle this case and avoid those specific micro-optimisations which are not as relevent as some other moves,
but the implementation of this mechanism still have some importance to converge faster.

\begin{figure}
\centering
\begin{tikzpicture}[scale=0.85]

% \node[gnode](o) at (0,0){};
% \node[gnode](x) at (4,0){};
% \node[gnode](y) at (0,4){};
% \node[gnode](p1) at (2,1){};
% \node[gnode](p2) at (1,2){};
% \draw[gedge](3,0) -- (0,3);
% \draw[gedge,dashed](p1) -- (o) -- (p2);
% \node[gnode](p1) at (2,1){};
% \node[gnode](p2) at (1,2){};
% \draw[gedge](x) -- (o) -- (y);
% \tkzMarkSegment[color=black,pos=.5,mark=||,size=2pt](u,v)
% \tkzMarkSegment[color=black,pos=.5,mark=||,size=2pt](v,(0,3))
% \tkzMarkSegment[color=black,pos=.5,mark=||,size=2pt](u,(3,0))

\node[gnode](o) at (0.0,0.0){};
\node[gnode](x) at (3.932,0.735){};
\node[gnode](y) at (-0.735,3.932){};
\draw[gedge](2.949,0.551) -- (-0.551,2.949);
\node[gnode](u) at (1.782,1.35){};
\node[gnode](v) at (0.615,2.15){};
\draw[gedge,dashed](u) -- (o) -- (v);
\draw[gedge](x) -- (o) -- (y);
\tkzMarkSegment[color=black,pos=.5,mark=||,size=2pt](u,v)
\tkzMarkSegment[color=black,pos=.5,mark=||,size=2pt](v,(-0.551,2.949))
\tkzMarkSegment[color=black,pos=.5,mark=||,size=2pt](u,(2.949,0.551))

% \node[draw=none](l1) at (1.782+0.2,1.35+0.2){$p_2$};
% \node[draw=none](l2) at (0.615+0.2,2.15+0.2){$p_1$};
% \node[draw=none](l3) at (0.2,-0.2){$u$};
% \node[draw=none](l4) at (3.932,0.735+0.2){$w_2$};
% \node[draw=none](l5) at (-0.735+0.3,3.932){$w_1$};

\foreach \lbl/\p/\s in {$u$/o/left,$w_1$/x/above,$w_2$/y/left,$p_1$/u/above right,$p_2$/v/above right}{
    \node[\s=0 of \p](){\lbl};
}
\end{tikzpicture}
\caption{Two possible oscillating positions $p_1$ and $p_2$ for a vertex $v$ if $\deltaE=0$ and $n_r=1$. The vertex $v$,
initially in position $p_1$ have for only neighbour the vertex $u$,
if the ray have the same slope as the segment $\overline{p_1 p_2}$ and hits the edge $(u,w_2)$,
the algorithm can evaluate the position $p_2$ as a valid new location. Once in $p_2$,
the same behavior can happen to evaluate $p_1$ as its new position.}
\label{fig:oscillating_positions}
\end{figure}
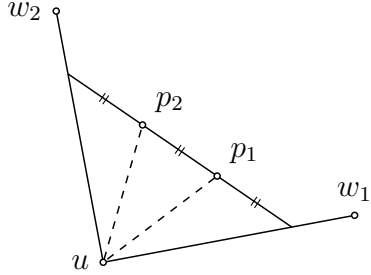

\subsection{Accessing facets} \label{subsection:accessing_facets}

The aim of this subsection is to find the number of reflections, or rays, needed in average to encounter a certain proportion $q\in[0,1]$ of the facets of the graph. Since we are interested in an applicable bound at any step of the algorithm, we will consider some of the worst cases in terms of graph configuration.

Let us begin with the following lemma:

\begin{lemma} \label{lm:proba_edge}
    Given $\GG=\structure{V,E}$ a graph and $\Gamma$ a drawing of $\GG$, a random semi-line intersects in average $\frac{|E|}{6}$ edges.
\end{lemma}

\begin{proof}
    Consider three points $A$, $B$ and $R$ chosen at random according to the uniform distribution on $[0,1]^2$. Among all the possible rays originating from $R$, the proportion of those which intersect the segment $AB$ (\ie those passing into the triangle $ABR$) equals $\frac{\widehat{ARB}}{2\pi}$. Furthermore, since for any three points, the sum of there three angles equals $\pi$, it is easy to convince oneself that the average angle formed by three points is $\frac{\pi}{3}$. This gives us the probability of $\frac{1}{6}$ that a random ray intersects a random segment. Note that this probability actually holds for drawings on any convex surface.

    Considering that the vertices of $\GG$ are uniformely distributed on $\boundingBox[\varepsilon]$, for any edge $e\in E$, a random ray has a probabilty of $\frac{1}{6}$ to cross it. Hence, it will intersects in average $\frac{|E|}{6}$ edges.

\end{proof}

Note that the consideration made here on the uniform distribution of the vertices is a rather strong assumption. In practice, as the algorithm progresses, the edges tends to be much smaller than if the vertices where placed randomly on $\boundingBox[\varepsilon]$. However, this whole reasoning will give us a limit for the start of the algorithm, which we will use all along. This allows us to prove the following:

\begin{theorem}\label{th:proba_facet}
    Given a graph $\GG=\structure{V,E}$ with $|V|$ and $|E|$ sufficiently large, and a drawing $\Gamma$ of $\GG$. An upper bound $P$ for the probability that $R$ random rays crosse a given facet equals:

    \[
    P=1-\left(1-\frac{|E|}{36(|V|+cr(\Gamma)-2)}\right)^{3R}
    \]

\end{theorem}

\begin{proof}
After the probability for a ray to cross an edge, one needs to have an idea of the number of facets in a generic graph. Considering that we know the number of crossings of our actual $\Gamma$, we can introduce an approximation for the number of facets. As explained before, the number of facets of a drawing $\Gamma$ is the number of faces its planarized version $\Gamma'$.
Given a graph $\GG=\structure{V,E}$ and a drawing $\Gamma$ of $\GG$ with, by definition, $cr(\Gamma)$ crossings, without more than two edges intersecting on the same point,
let $\GG'=\structure{V',E'}$ be the graph obtained from the planarization $\Gamma'$ of $\Gamma$.
We then must have $|V'|=|V|+cr(\Gamma)$ and $|E'|=|E|+2\cdot cr(\Gamma)$.
From the definition of the planarization, the new number of vertices is self-evident. For the number of edges, we can see that each crossing between $2$ edges leads to $4$ new edges in $\GG'$.

Now to count the number of faces of $\Gamma'$, we can, as we said, consider the worst case, namely, if all the faces are triangular. In this specific case, we can express the number of edges according to the number of faces $|E|=\frac{3F}{2}$, with F the number of faces. Moreover, thanks to Euler's formula for planar graphs, we know that $|V|-|E|+F=2$, which naturally leads to $F=2|V|-4$.

If we consider again that these facets are all triangular, meaning that all of the $2(|V|+cr(\Gamma))-4$ facets of $\GG$ have $3$ associate edges, the edges must have in average $F=\frac{6(|V|+cr(\Gamma)-2)}{|E|}$ joint facets.

To have the probability $Q$ that one specific facet is traversed by a random ray, we multiply the probability that, among the $3$ possible attached edges, $k$ are crossed, by the probability that, for at least one of them, the ray enters the good facet after having crossed it. This give us the following probability $Q$:
\[
Q=\sum_{k=0}^3 \binom{3}{k} \left(\frac{1}{6}\right)^k \left(\frac{5}{6}\right)^{3-k} \left(1-\left(1-\frac{1}{F}\right)^k\right)
\]

We consider $E$ and $V$ sufficiently large to have independent events and apply \emph{Bernoulli trials}.

We can then rearrange the terms to have:

\begin{align*}
    Q &= \sum_{k=0}^3 \binom{3}{k} \left(\frac{1}{6}\right)^k \left(\frac{5}{6}\right)^{3-k} - \sum_{k=0}^3 \binom{3}{k} \left(\frac{1}{6}\right)^k \left(\frac{5}{6}\right)^{3-k}\left(1-\frac{1}{F}\right)^k \\
    &= \left( \frac{1}{6}+\frac{5}{6} \right)^3 - \left(\frac{1}{6}\left(1-\frac{1}{F}\right) + \frac{5}{6} \right)^3 \\
    &= 1-\left(1-\frac{1}{6F}\right)^{3}
\end{align*}

This ending result can be interpreted as not hitting the good edge nor entering the good facet three times.

To finally have the probability $P$ for one facet to be hit by at least one of $R$ rays, we can apply the same process and have $P=1-(1-Q)^R$, which gives us the probability stated in the theorem.

\end{proof}

If we want that the proportion $q$ of facets that are encountered, then we need to find the $R$ such that $P\geq q$. To have a real idea about the number of rays, for all the graphs of our database on which we applied a random layout, reaching a $q\leq0.1$ would require around $10$ rays.

% !TEX root = ./main.tex

%
\section{Convergence and Complexity}\label{sec:properties}
%

% \begin{lemma}
%     Given $\GG=\structure{V,E}$ a planar graph and $\Gamma_0$ a convex drawing of $\GG$, if a vertex $v \in V$, not lying on the external face, is moved to another position $p$ in $\boundingBox[\varepsilon]$. Any ray cast with an irrational angle from $v$ will give, with a sufficient number of steps, a new position $p \prime$ included in the meta-face of $v$ in $\Gamma_0$.
% \end{lemma}
%
% \begin{proof}
%
% \end{proof}

This section gives some rather important properties of the algorithm, namely its convergence and its complexity.

First of all, we will give some lemmas helping to prove the convergence theorem.

\begin{lemma}\label{lemma:energy_bounded}
    Given a graph $\GG$ and an initial drawing $\Gamma_0$, $\exists \energy_{MAX}$ such as $\forall n \in \N$, $\energy[\Gamma_n]\leq\energy_{MAX}$.
    %At any point of the algorithm, the energy of $\Gamma$ is bounded.
\end{lemma}

\begin{proof}
Since the frame inside which the vertices can move is fixed at the beginning of the algorithm and never expand during the next steps,
the maximum length of an edge $e$ is bounded by the diagonal of this frame.
As a result, the energy of one edge $e$, since it simply depends quadratically on its length, is also bounded. The maximum energy is either if both of its endpoints are in opposite corners or if they are at the same place.
In addition, since the number of edges is also bounded,
the total energy of $\Gamma$ is bounded by a hypothetical value $\energy_{MAX}$.

\end{proof}

\begin{lemma}\label{lemma:local_equals_global}
    If a vertex $v$ is moved to a place which minimizes its local energy, the global energy of $\Gamma$, i.e. $\energy[\Gamma]$, can only decrease.
\end{lemma}

\begin{proof}

The local energy of $v$ is defined by $\energy[v]=\sum_{e \in E_v}{\energy[e]}$. When we move the vertex $v$ to a another location, the energy of the edges non-attached to $v$, $\overline{\energy}(v)=\sum_{e \in E \backslash E_v}{\energy[e]}$ is unchanged. Since $\energy[\Gamma]=\energy[v]+\overline{\energy}(v)$, if $\energy[v]$ is decreased by the repositioning of $v$, $\energy[\Gamma]$ can only decrease too.

\end{proof}

\begin{lemma}\label{lemma:monotonic_decrease_energy}
    With $\deltaE>0$, when we move only one vertex $v$, there exists a number of steps $n_s$ after which either $v$ is blocked in a local minima or a place which decrease the number of crossings of $v$ is found.
\end{lemma}

\begin{proof}

Since the energy of $\Gamma$ is bounded and the use of $\deltaE$ makes it decrease by fixed quantified steps. The number of steps, before reaching an energy of $0$ must be finite. We could even determine that its value equals $n_s = \ceil*{\ffrac{\energy_{MAX}}{\deltaE}}$.

\end{proof}

We can now propose the following convergence theorem.

\begin{lemma}\label{lemma:monotonic_decrease_crossing}
    Given a graph $\GG$ and an initial drawing $\Gamma_0$, $\forall n\in \N$, $cr(\Gamma_n)\geq cr(\Gamma_{n+1})$.
\end{lemma}

\begin{proof}
By calling the $\texttt{move}(\texttt{v},\Gamma)$ function,
the position that it returns can not have a worse crossing number than the initial position $p$ of $v$.
Since the first criterion of comparison,
if every potential positions obtained by the rays have a higher crossing number, $p$ is returned.
Thus, after each iteration of the main loop in the \texttt{RRGD} function, the crossing number has decreased or remained stationary.

\end{proof}

\begin{theorem}\label{th:convergence}
Given a graph \GG, and an initial drawing $\Gamma_0=\Gamma$, the algorithm always converges and stops.
\end{theorem}

\begin{proof}
For each application of the global movement function, one of this outcome can appear:
\begin{enumerate}
    \item either $cr(\Gamma)$ remains the same, and $\energy[\Gamma]$ decreases \label{proof1:enum1}
    \item or $cr(\Gamma)$ decreases and $\energy[\Gamma]$ is set to another value but bounded by $\energy_{MAX}$ \label{proof1:enum2}
\end{enumerate}

First, the case \ref{proof1:enum1} can only appear a finite number of times before we are forced to enter case \ref{proof1:enum2} or to stop completely the algorithm since the decrease of the energy is discretized by the energy delta $\deltaE$, as shown by Lemmas~\ref{lemma:monotonic_decrease_energy} and~\ref{lemma:local_equals_global}.
Second, the number of times we go into case~\ref{proof1:enum2} is also finite: not only the crossing number can not increase by Lemma~\ref{lemma:monotonic_decrease_crossing}, but
once it reach the theoretical crossing number of \GG, it can not go under (obviously, we will almost always stop before).
Moreover, since $\energy[\Gamma]$ is bounded following Lemma~\ref{lemma:energy_bounded},
we also can not start an infinite loop of case~\ref{proof1:enum1}.

\end{proof}

\begin{proposition}
    Given a graph $\GG=\structure{V,E}$ and one drawing $\Gamma_n$, $\Gamma_{n+1}$ can be computed in $O(|E|^2Rn_r)$ with $R$ the number of rays cast for each vertex and $n_r$ their sizes.
\end{proposition}

\begin{proof}

To move one node $v$, we thus cast $R$ rays, with each one of them reflecting or crossing $n_r$ times.
Since the computations of each intersection between a ray and the edges is in $O(|E|)$ and the computation of the opacity depends on the degree of $v$ (which we will denote $k$).
The complexity of the \texttt{move} function is in $O(k|E|Rn_r)$. Calling this for at most each node in $\GG$ leads to a complexity of $O(|E|^2Rn_r)$ to go from a $\Gamma_n$ to $\Gamma_{n+1}$.

\end{proof}

% !TEX root = ./main.tex

%
\section{Performances and experiments}\label{sec:perfs}

\subsection{Testing process}

In order to assess the performances of our algorithm, we tested it on instances of graphs present in the dataset of the graph drawing community~\footnote{Graph drawing benchmarks from \texttt{graphdrawing.org} can be found at \texttt{http://www.graphdrawing.org/data.html}} and also with random 3-connected graphs constructed by starting from $K_4$ and iteratively inserting new edges between the middle of two previous ones.
The final experiments (Figure~\ref{fig:crossings_and_time}) have been done on 500 graphs of each of the main classes in this dataset, mainly, \emph{NORTH}, \emph{ROME} and \emph{DAG}. We compared our results with the best Rectilinear Crossing Minimization one of our knowledge, \ie \texttt{StressMinimization}, implemented in \textit{OGDF}. Note that \texttt{StressMinimization} has been chosen over \texttt{SpringEmbedderKK} since it produces in average fewer crossings.
However, before comparing frontally with \texttt{StressMinimization}, we will investigate first the empiric influence of our parameters on the results. Note that all the results shown are within the \emph{three-sigma limit} (\ie values further from the mean by more than three times the standard deviation will be ommitted).

Note that we ran all the experiments on an Intel Core i7-10850H processor running at 2.71GHz, with 16GB of RAM. The algorithm has been implemented in Python 3.7.10 and run with the version 7.3.5 of PyPy.

\subsection{Drawing initialization} \label{sec:drawingInit}

\begin{figure}
\begin{center}
\includegraphics[width=\textwidth]{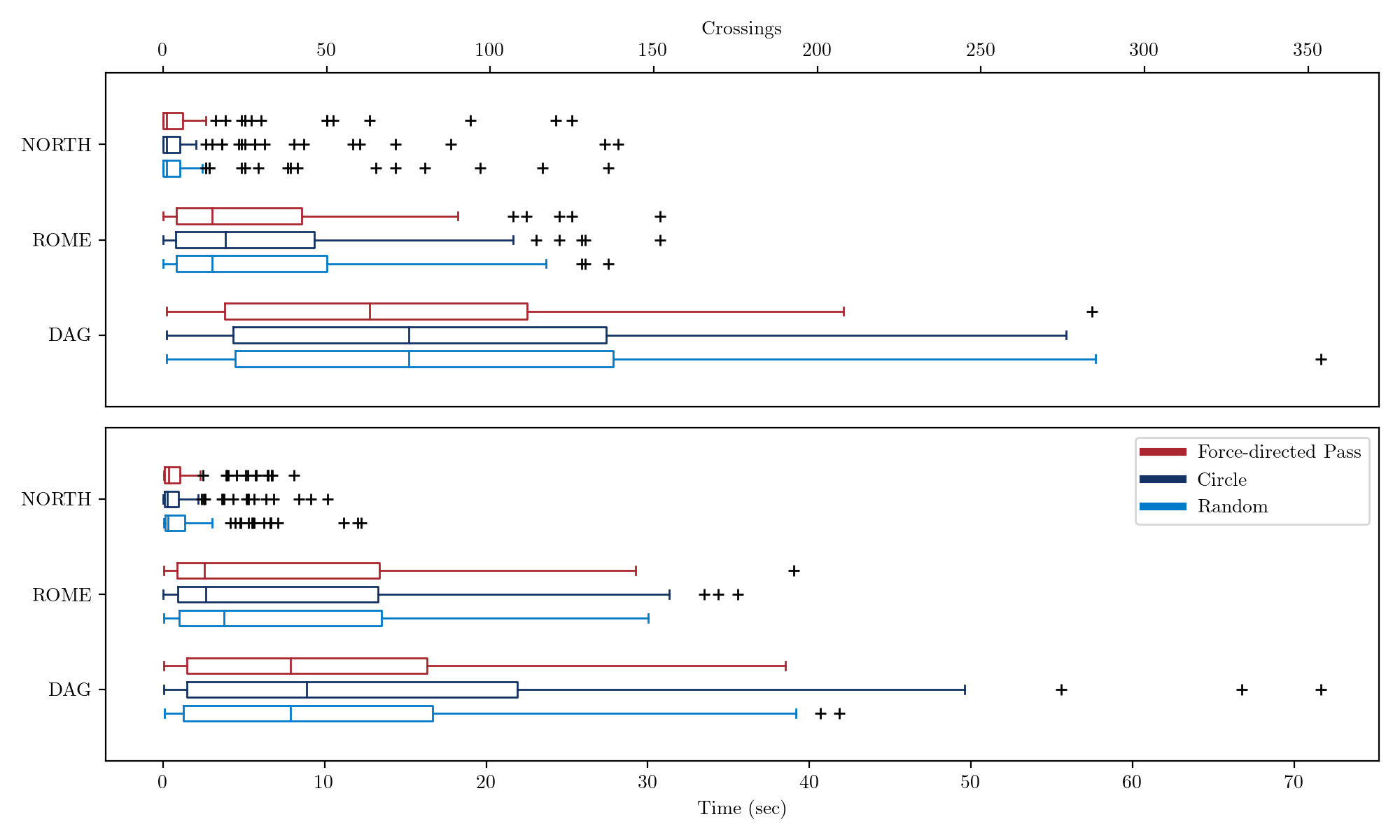}
\caption{Plots showing for the three classes of graphs and for the different initialization layouts, the resulting number of crossings (left) and the computation time (right).}
\label{fig:drawing_init}
\end{center}
\end{figure}

Our algorithm needs an existing drawing to perform. In the basic behavior,
the x and y coordinates of the vertices are initialized randomly in preset intervals set to respectively the width and the height of the visualization window.
We tried other simple initial configurations such as the circle layout
(where vertices are trivially placed uniformly on the boundary on a circle according to there indices).
We also considered a quick pass with a custom force-directed layout (this pass is just few iterations implementing only Hooke's and Coulomb's, nothing comparable to the \texttt{StressMinimization} algorithm that we speaked about).
However, the goal of the initial layout is only to quicken the total computation time of the algorithm, we must ensure that the initialization is not too ``powerful" for our algorithm, meaning that the final result does not depends on it.
An overview of the influence of these layouts are shown in Figure \ref{fig:drawing_init}.
In addition, to make sure of it, we made statistical tests (\ie \emph{Two-sample t-Test} with a threshold of $1\%$) to verify the hypotheses which say that the average crossing number is the same for every starting layout. None of these hypotheses have been rejected.

\subsection{Energy delta}

The \deltaE provides an equilibrium between the computation time and the refinement of the solution.
The smaller it is, the more the algorithm is careful to move the vertices.
On the other hand, with a bigger \deltaE,
the algorithm do not bother to move vertices bit by bit and can potentially pass to other vertices for which their displacement is more impactful.
The aim here is to find a good balance between these two aspects.

\subsection{Prohibition window}

The prohibition window \prohibitionWindow has a similar role to the \deltaE in the behavior of the algorithm,
but in a more direct way.
The smaller it is, the more the algorithm spends time on specific vertices.
With a large \prohibitionWindow,
the vertices move more frequently which allows to find a stable state much quicker.
Note that a too large \prohibitionWindow affects badly the process,
diminishing the amount of vertices that can be evaluated in an iteration.
Eventually, a $\prohibitionWindow \geq |V|$ allows the vertices to move only once, which gives obviously some rather bad results.

\begin{figure}
\centering
\begin{minipage}{.5\textwidth}
  \centering
  \includegraphics[width=\linewidth]{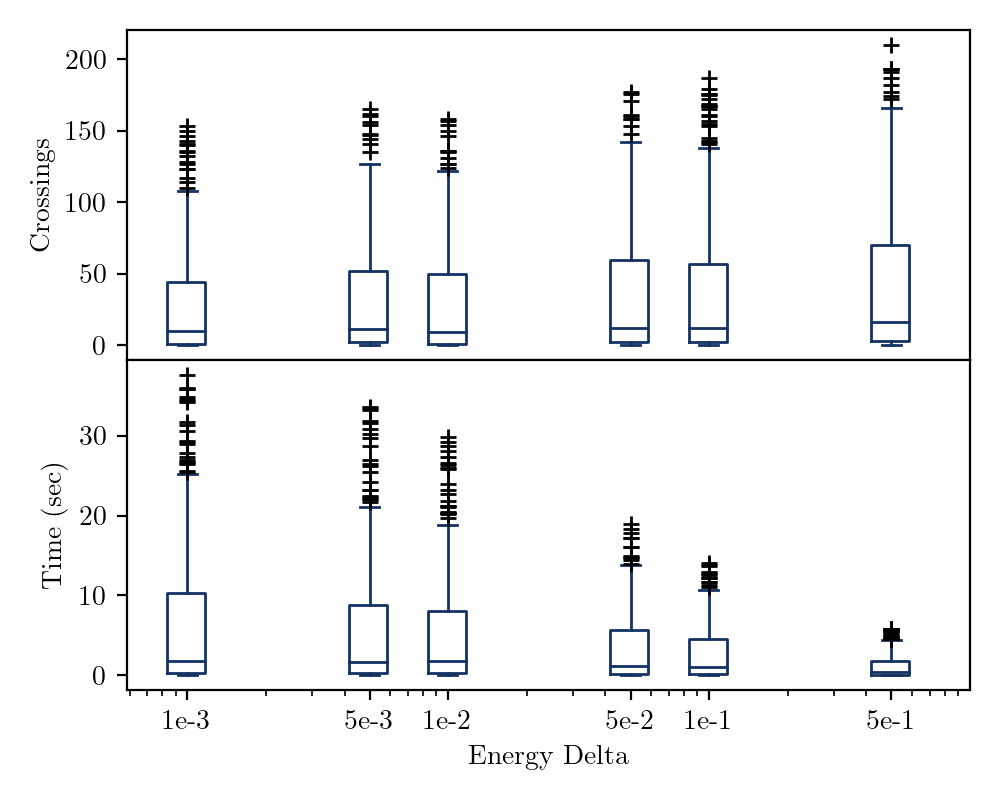}
  \label{fig:ed}
\end{minipage}%
\begin{minipage}{.5\textwidth}
  \centering
  \includegraphics[width=\linewidth]{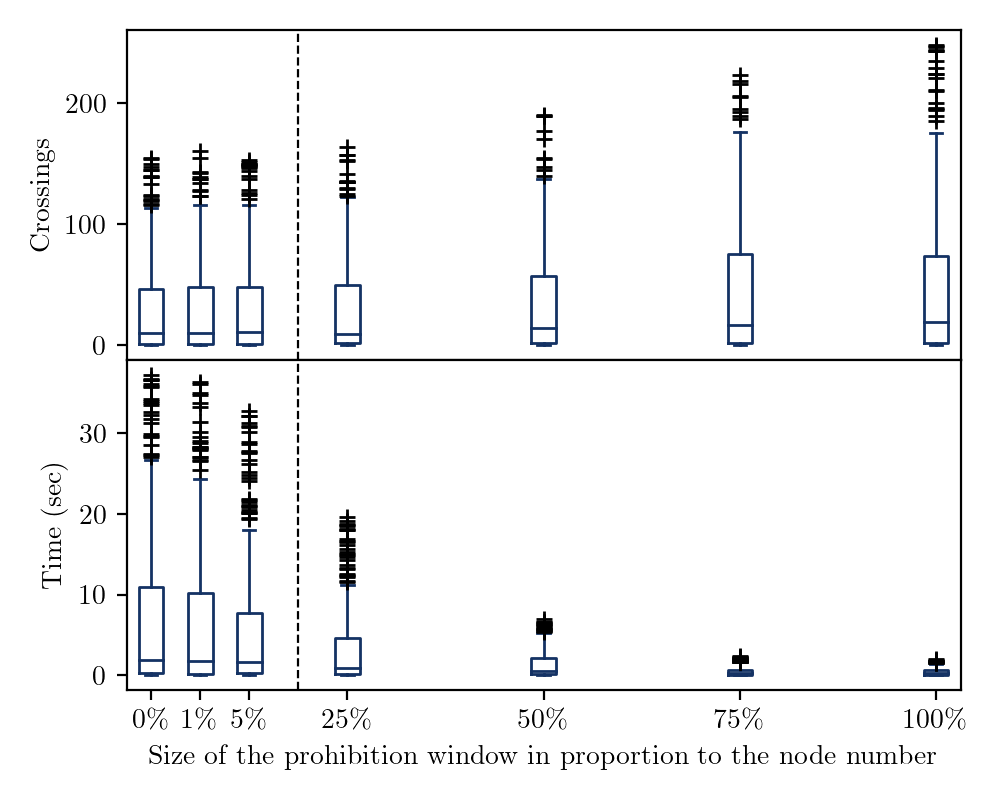}
  \label{fig:pw}
\end{minipage}
\caption{Plots showing the number of crossings (tops) and the execution time (bottoms) according to the energy delta (left) and the length of the prohibition window expressed as a ratio of the node number of the graphs (right).}
\label{fig:ed_and_pw}
\end{figure}

Empirically, these consideration can be seen. Figure~\ref{fig:ed_and_pw} shows the inverse proportionality of the number of crossings and the time needed. Note that in this figure, the size of the prohibition window is in proportion to the total node number.

\subsection{Crossing behavior}

\begin{figure}
\centering
\begin{minipage}{.85\textwidth}
  \centering
  \includegraphics[width=\linewidth]{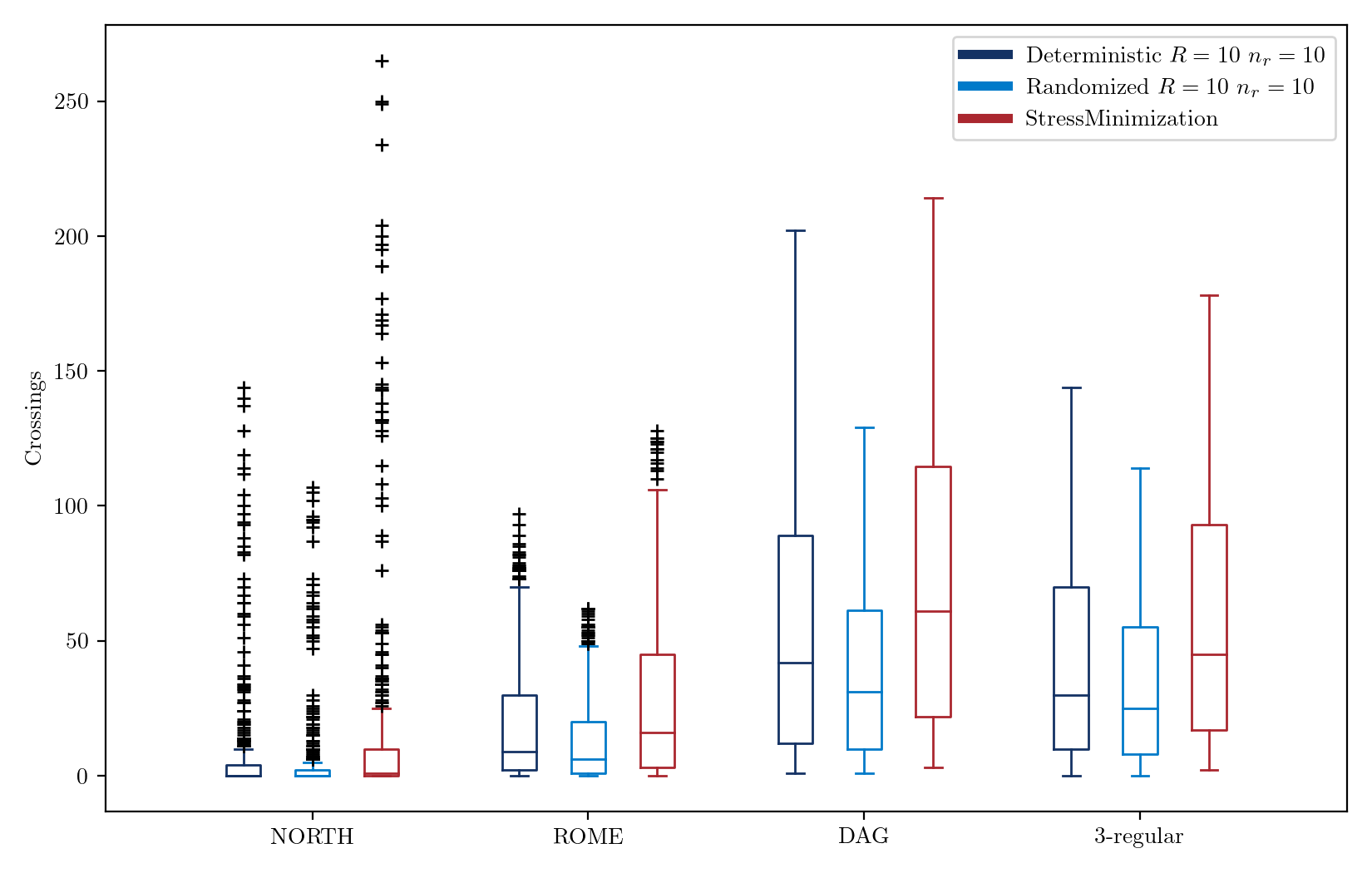}
  \label{fig:crossings}
\end{minipage}

\begin{minipage}{.85\textwidth}
  \centering
  \includegraphics[width=\linewidth]{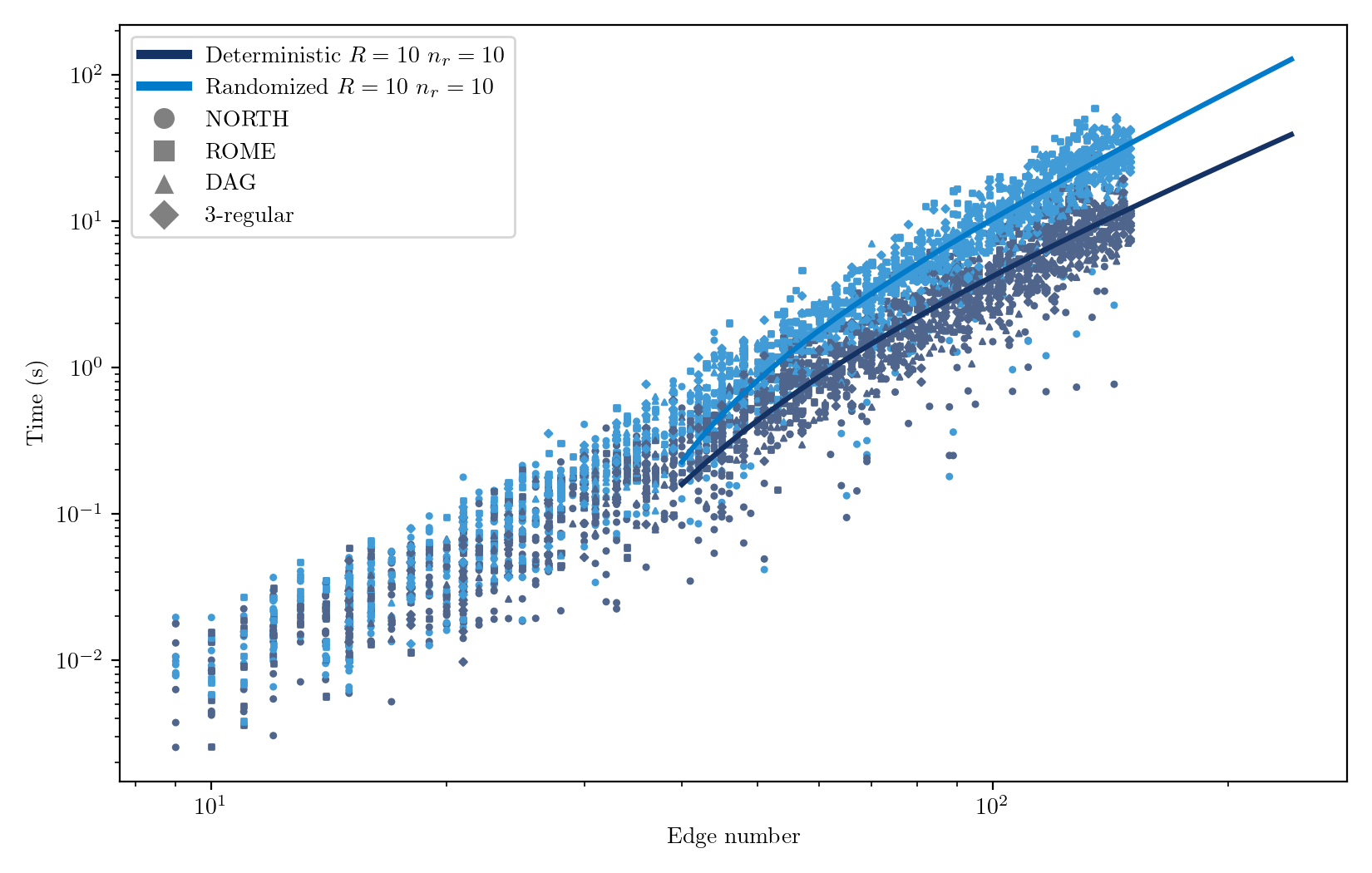}
  \label{fig:time}
\end{minipage}
\caption{Plots showing for the four classes of graphs and for the different algorithms: deterministic, randomized (refering to how the opacity is considered) with $R=10$ and $n_r=10$ for each of these and \texttt{StressMinimization}, the resulting number of crossings (left) and the computation time according to the number of edges of the considered graph (right)}
\label{fig:crossings_and_time}
\end{figure}

The influence of the opacity mode is more subtle than the energy delta and the prohibition window. Deterministic crossings tend to consolidate satisfying configurations of $\Gamma$. On the other hand, randomized ones allow, as we said, to go beyond a locally worse choice to find a better one after. This results in having fewer edge crossings overall, as we see in the Figure~\ref{fig:crossings_and_time}.

% \begin{figure}
% \begin{center}
% \includegraphics[width=.6\textwidth]{pics/crossings.png}
% \caption{Plots showing for the three classes of graphs and for the three different algorithms, the resulting number of crossings.}
% \label{fig:crossings}
% \end{center}
% \end{figure}

In the latter, we compared the crossings obtained from our two versions of opacity behavior with the ones obtained from the \texttt{StressMinimization} algorithm. Following Section~\ref{subsection:accessing_facets}, we chose $R=10$ to enter in average a tenth of the facets of the graph under consideration. In addition, we chose also $n_r=10$, to potentially enter all the facets (considering that the rays do not reflect too early and that two rays do not enter the same facet).

The two versions have significantly better results than the existing algorithm on \emph{OGDF}. Moreover, the number of crossings from the randomized approach is lower, as expected, than the deterministic one.

% \begin{figure}
% \begin{center}
% \includegraphics[width=.6\textwidth]{pics/time.png}
% \caption{Plots showing for the three classes of graphs and our two different algorithms, the computation time according to the number of edges of the considered graph.}
% \label{fig:time}
% \end{center}
% \end{figure}

These results come at a certain cost: the computation time. Figure~\ref{fig:crossings_and_time} shows this execution time for our two version. The time is expressed as a function of the number of edges of the graph, since this is the main parameter in the complexity of our algorithm. Again, the results are in correlation with our expectations, the deterministic one converges faster than the randomized one. This is simply due to the fact that the consideration of the opacity only as a bias allow to go beyond some local minima, and push back the convergence of the system to find better solutions.
Note that the figure also shows that total convergence time is clearly polynomial.

\begin{figure}
\centering
\begin{minipage}{.85\textwidth}
  \centering
  \includegraphics[width=\linewidth]{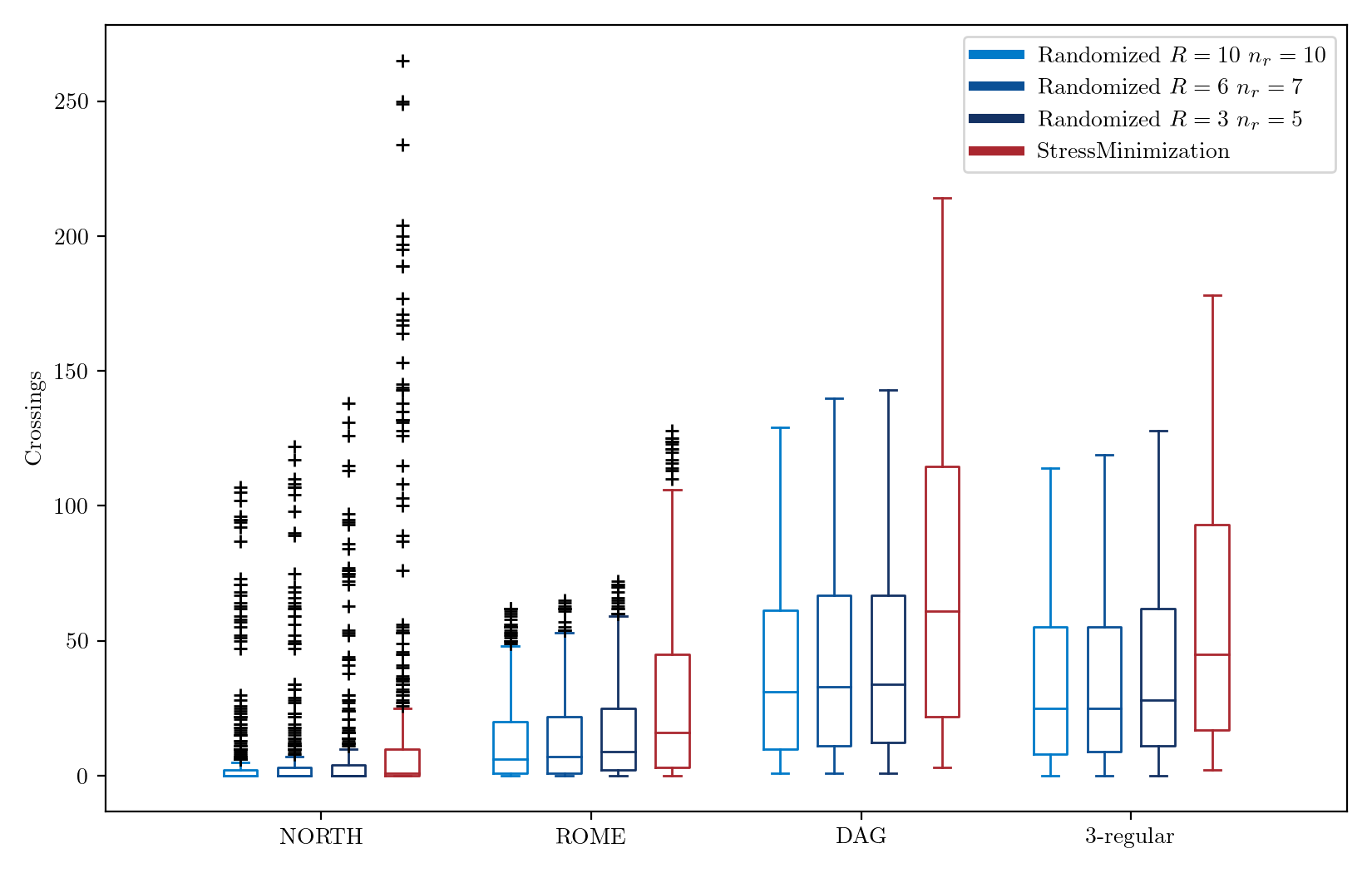}
  \label{fig:crossings_all}
\end{minipage}

\begin{minipage}{.85\textwidth}
  \centering
  \includegraphics[width=\linewidth]{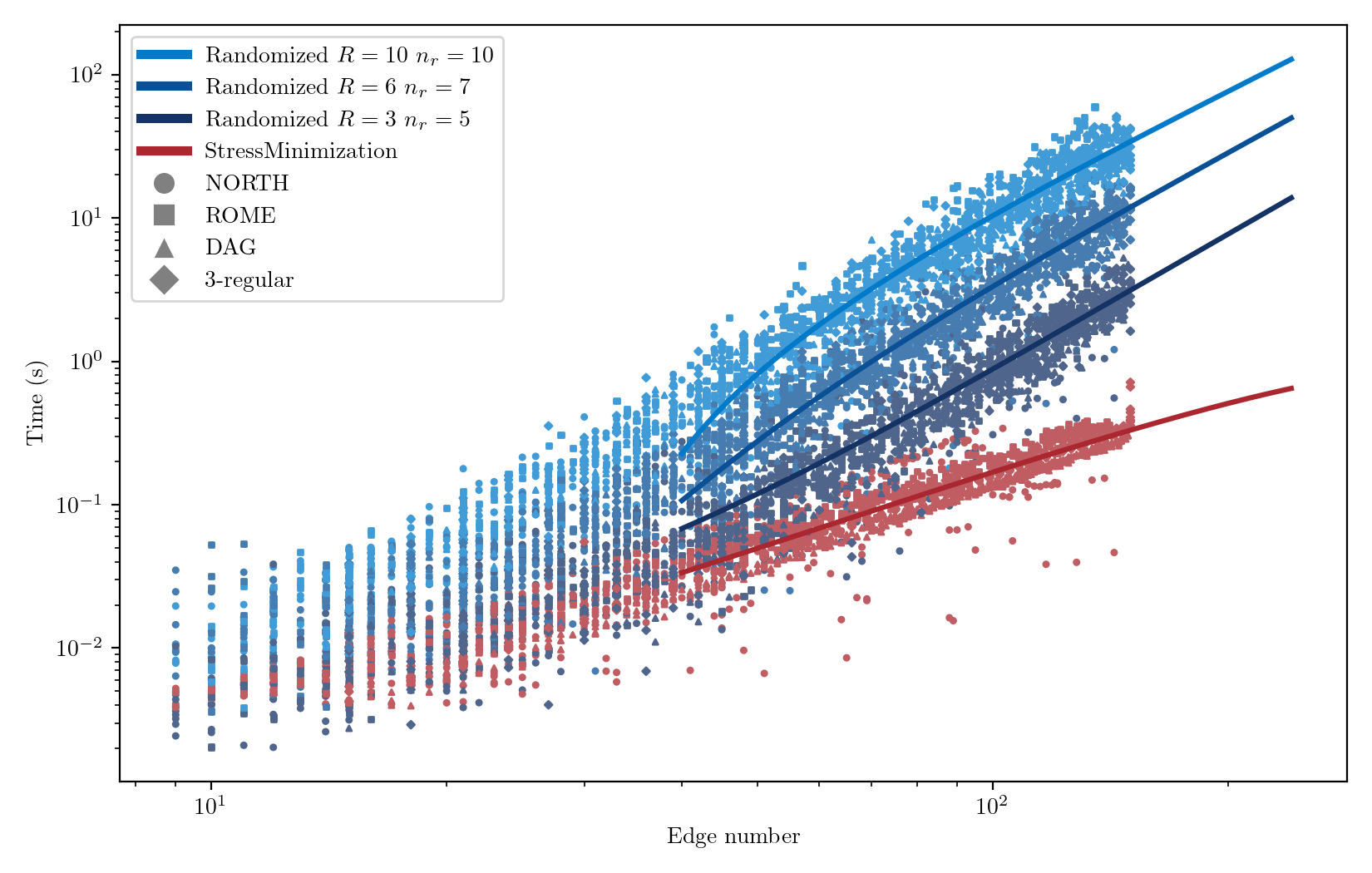}
  \label{fig:time_all}
\end{minipage}
\caption{Plots showing for the four classes of graphs and for different values of $R$ and $n_r$, the resulting number of crossings (left) and the computation time according to the number of edges of the considered graph (right), again compared with \texttt{StressMinimization}}
\label{fig:crossings_and_time_all}
\end{figure}

Finally, by reducing the number parameters $R$ and $n_r$, the resultings crossings do not deteriorate that much. However, the computation time, strongly depending on these values, decrease sharply to go around the second for the bigger graphs, always under \texttt{StressMinimization} but within a reasonable time.

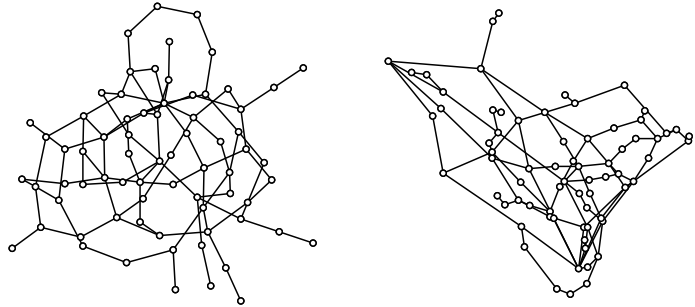
\begin{figure}
\centering
\begin{tikzpicture}
\begin{scope}[xshift=-5cm]
\foreach \i/\x/\y in {0/1.70/1.62,1/0.13/1.66,2/1.70/1.09,3/0.70/1.60,4/1.39/1.15,5/4.00/0.81,6/1.74/1.40,7/2.52/0.78,8/2.41/2.48,9/2.11/1.98,10/2.64/0.24,11/0.70/2.06,12/2.46/3.85,13/2.09/3.49,14/0.45/2.22,15/1.93/3.96,16/1.45/2.79,17/1.19/2.81,18/1.96/1.90,19/2.02/2.67,20/0.94/2.03,21/0.94/1.59,22/0.95/2.50,23/3.11/2.06,24/2.55/1.81,25/3.04/2.59,26/2.14/1.59,27/2.40/2.78,28/2.54/1.28,29/3.04/0.04,30/2.60/0.96,31/3.13/1.35,32/3.35/1.88,33/2.84/0.48,34/3.48/2.88,35/2.89/1.75,36/0.00/0.74,37/1.22/2.22,38/2.08/2.98,39/0.31/1.56,40/2.90/1.47,41/1.75/2.54,42/2.32/2.27,43/3.55/0.96,44/1.96/0.91,45/2.14/0.72,46/2.46/1.42,47/1.53/2.46,48/3.01/2.29,49/1.57/3.09,50/0.62/1.38,51/1.47/1.62,52/2.57/2.79,53/3.45/1.65,54/1.54/3.60,55/0.94/0.77,56/3.87/3.14,57/3.04/1.13,58/1.90/3.13,59/0.91/0.89,60/2.17/0.19,61/1.93/2.52,62/1.52/0.55,63/2.66/3.35,64/0.24/2.41,65/1.55/2.25,66/0.38/1.02,67/2.87/2.86,68/1.55/2.00,69/2.77/2.16,70/1.23/1.68}{
\node[gnode](\i) at (\x,\y){};
}
\foreach \u/\v in {0/2,0/18,0/26,0/68,0/70,1/3,1/39,2/44,3/70,4/6,4/44,4/59,4/70,5/43,6/9,7/10,7/46,8/9,8/19,8/67,8/69,11/50,11/37,11/64,12/15,12/63,13/38,14/22,14/39,15/54,16/17,16/19,16/22,16/49,17/65,18/46,18/51,18/61,18/65,19/38,19/42,19/47,19/41,19/58,19/52,20/21,20/22,20/70,21/51,22/37,23/24,23/25,23/53,24/26,24/28,24/42,25/27,25/34,25/67,27/41,28/30,28/45,28/35,29/33,30/31,30/33,30/44,31/32,31/53,31/57,32/48,34/56,35/40,35/48,35/69,36/66,37/41,37/47,37/70,38/61,39/50,39/66,40/46,41/61,43/57,45/60,45/62,46/57,47/68,48/52,49/54,49/61,49/58,50/55,52/63,55/62,59/66}{
\draw[gedge] (\u) -- (\v);
}
\foreach \i/\x/\y in {0/1.70/1.62,1/0.13/1.66,2/1.70/1.09,3/0.70/1.60,4/1.39/1.15,5/4.00/0.81,6/1.74/1.40,7/2.52/0.78,8/2.41/2.48,9/2.11/1.98,10/2.64/0.24,11/0.70/2.06,12/2.46/3.85,13/2.09/3.49,14/0.45/2.22,15/1.93/3.96,16/1.45/2.79,17/1.19/2.81,18/1.96/1.90,19/2.02/2.67,20/0.94/2.03,21/0.94/1.59,22/0.95/2.50,23/3.11/2.06,24/2.55/1.81,25/3.04/2.59,26/2.14/1.59,27/2.40/2.78,28/2.54/1.28,29/3.04/0.04,30/2.60/0.96,31/3.13/1.35,32/3.35/1.88,33/2.84/0.48,34/3.48/2.88,35/2.89/1.75,36/0.00/0.74,37/1.22/2.22,38/2.08/2.98,39/0.31/1.56,40/2.90/1.47,41/1.75/2.54,42/2.32/2.27,43/3.55/0.96,44/1.96/0.91,45/2.14/0.72,46/2.46/1.42,47/1.53/2.46,48/3.01/2.29,49/1.57/3.09,50/0.62/1.38,51/1.47/1.62,52/2.57/2.79,53/3.45/1.65,54/1.54/3.60,55/0.94/0.77,56/3.87/3.14,57/3.04/1.13,58/1.90/3.13,59/0.91/0.89,60/2.17/0.19,61/1.93/2.52,62/1.52/0.55,63/2.66/3.35,64/0.24/2.41,65/1.55/2.25,66/0.38/1.02,67/2.87/2.86,68/1.55/2.00,69/2.77/2.16,70/1.23/1.68}{
\node[gnode](\i) at (\x,\y){};
}
\end{scope}
\begin{scope}[xshift=0cm]
\foreach \i/\x/\y in {0/2.53/1.83,1/3.29/2.20,2/2.33/2.16,3/3.11/2.06,4/2.65/2.20,5/0.31/3.08,6/2.46/1.90,7/1.39/2.58,8/2.15/1.23,9/2.28/1.56,10/1.50/2.55,11/3.99/2.17,12/2.24/0.20,13/2.61/0.85,14/3.46/1.94,15/2.42/0.13,16/2.85/1.10,17/2.64/1.28,18/2.34/1.63,19/2.53/0.47,20/3.06/1.73,21/2.89/1.69,22/3.26/1.63,23/1.38/1.94,24/1.87/1.80,25/1.73/1.38,26/2.22/1.83,27/2.15/1.16,28/1.73/2.44,29/1.47/3.87,30/1.23/3.13,31/0.00/3.23,32/0.59/2.50,33/1.39/3.76,34/1.55/1.32,35/1.38/2.02,36/3.41/2.57,37/3.15/1.55,38/2.62/0.74,39/3.58/2.20,40/1.34/2.14,41/2.60/1.02,42/2.20/1.14,43/0.51/3.05,44/2.08/2.55,45/2.48/2.67,46/1.46/2.28,47/2.83/1.14,48/0.73/1.74,49/2.79/0.63,50/3.83/2.32,51/2.67/1.65,52/1.77/1.03,53/0.70/2.60,54/2.64/0.27,55/3.70/2.27,56/1.45/1.44,57/0.73/2.82,58/2.65/0.49,59/3.00/2.34,60/2.35/2.78,61/2.65/1.02,62/3.14/2.91,63/1.81/0.76,64/4.00/2.23,65/2.54/1.47,66/3.36/2.45,67/1.88/1.31,68/2.68/1.47,69/1.76/1.62,70/2.93/1.87}{
\node[gnode](\i) at (\x,\y){};
}
\foreach \u/\v in {0/2,0/18,0/26,0/68,0/70,1/3,1/39,2/44,3/70,4/6,4/44,4/59,4/70,5/43,6/9,7/10,7/46,8/9,8/19,8/67,8/69,11/50,11/37,11/64,12/15,12/63,13/38,14/22,14/39,15/54,16/17,16/19,16/22,16/49,17/65,18/46,18/51,18/61,18/65,19/38,19/42,19/47,19/41,19/58,19/52,20/21,20/22,20/70,21/51,22/37,23/24,23/25,23/53,24/26,24/28,24/42,25/27,25/34,25/67,27/41,28/30,28/45,28/35,29/33,30/31,30/33,30/44,31/32,31/53,31/57,32/48,34/56,35/40,35/48,35/69,36/66,37/41,37/47,37/70,38/61,39/50,39/66,40/46,41/61,43/57,45/60,45/62,46/57,47/68,48/52,49/54,49/61,49/58,50/55,52/63,55/62,59/66}{
\draw[gedge] (\u) -- (\v);
}
\foreach \i/\x/\y in {0/2.53/1.83,1/3.29/2.20,2/2.33/2.16,3/3.11/2.06,4/2.65/2.20,5/0.31/3.08,6/2.46/1.90,7/1.39/2.58,8/2.15/1.23,9/2.28/1.56,10/1.50/2.55,11/3.99/2.17,12/2.24/0.20,13/2.61/0.85,14/3.46/1.94,15/2.42/0.13,16/2.85/1.10,17/2.64/1.28,18/2.34/1.63,19/2.53/0.47,20/3.06/1.73,21/2.89/1.69,22/3.26/1.63,23/1.38/1.94,24/1.87/1.80,25/1.73/1.38,26/2.22/1.83,27/2.15/1.16,28/1.73/2.44,29/1.47/3.87,30/1.23/3.13,31/0.00/3.23,32/0.59/2.50,33/1.39/3.76,34/1.55/1.32,35/1.38/2.02,36/3.41/2.57,37/3.15/1.55,38/2.62/0.74,39/3.58/2.20,40/1.34/2.14,41/2.60/1.02,42/2.20/1.14,43/0.51/3.05,44/2.08/2.55,45/2.48/2.67,46/1.46/2.28,47/2.83/1.14,48/0.73/1.74,49/2.79/0.63,50/3.83/2.32,51/2.67/1.65,52/1.77/1.03,53/0.70/2.60,54/2.64/0.27,55/3.70/2.27,56/1.45/1.44,57/0.73/2.82,58/2.65/0.49,59/3.00/2.34,60/2.35/2.78,61/2.65/1.02,62/3.14/2.91,63/1.81/0.76,64/4.00/2.23,65/2.54/1.47,66/3.36/2.45,67/1.88/1.31,68/2.68/1.47,69/1.76/1.62,70/2.93/1.87}{
\node[gnode](\i) at (\x,\y){};
}
\end{scope}

\end{tikzpicture}
\caption{Representation of \emph{grafo9660.71.graphml} in \emph{ROME} class \emph{StressMinimization} (left) and \emph{RBGD} (right) and with respectively $53$ and $21$ crossings.}
% \caption{Representation of \emph{grafo9660.71.graphml} in \emph{ROME} class with \texttt{SpringEmbedderKK} (left), \texttt{StressMinimization} (middle) and \texttt{RBGD} (left) and with respectively $47$,$53$ and $21$ crossings.}
\end{figure}

\subsection{Sources}

For testing purposes, full sources are available upon request. 

% !TEX root = ./main.tex

\section{Conclusion}
In this paper we proposed a new heuristic for the \emph{Rectilinear Crossing Minimization Problem} based on the basic principle of iteratively moving vertices along the plane.

%We studied rectilinear concepts and proposed a heuristic for the \emph{Rectilinear Crossing Minimization Problem} based on the basic principle of iteratively moving vertices along the plan.

%Thanks to a new approach in the moving mechanism relying on casting rays that reflects on the edges of the graph, we introduced an algorithm which have a competitive complexity compared to the other vertex-moving algorithm.

The main novelty consists in the moving mechanism which is based on an idea of casting rays against the edges of the graph. The new algorithm has a competitive complexity compared to the other vertex-moving algorithms.

We also discussed various ways to tune the algorithm parameters and see how to trade precision for time. We studied some geometrical properties connected to our algorithm in order to avoid the behaviors that can lead to edge cases.
Benchmarks show that the proposed algorithm causes fewer crossings than the best competitor (at the best of our knowledge) whose implementation is available.

Several improvements of the algorithm are possible: the first one being the parallelization of the code to speed up the execution.
A second option would be to introduce better data-structures to improve the complexity of finding the intersections between the edges and a ray.
Currently, even though it is not fully detailled in the paper, we already optimized this step by sorting the edges by their left endpoint and applying a dichotomic search on them but a clever algorithmic trick could possibly perform better.
Finally, it would be interesting to compare our algorithm with other recent ones based on the same principle. Unfortunetly, we should reimplement one based on a paper, since no algorithm of this kind is available in \emph{OGDF}, and no code is available elsewhere.

\bibliography{main}

\begin{thebibliography}{10}
\expandafter\ifx\csname url\endcsname\relax
  \def\url#1{\texttt{#1}}\fi
\expandafter\ifx\csname urlprefix\endcsname\relax\def\urlprefix{URL }\fi
\expandafter\ifx\csname href\endcsname\relax
  \def\href#1#2{#2} \def\path#1{#1}\fi

\bibitem{OGDF}
M.~Chimani, C.~Gutwenger, M.~J{\"u}nger, G.~W. Klau, K.~Klein, P.~Mutzel, {T}he
  {O}pen {G}raph {D}rawing {F}ramework ({OGDF}), in: R.~Tamassia (Ed.),
  Handbook of Graph Drawing and Visualization, CRC Press, 2012, pp. 543--570.

\bibitem{tutte-1963}
W.~T. Tutte, How to draw a graph, Proceedings of the London Mathematical
  Society S3-13~(1) (1963) 743--767.

\bibitem{spring_and_force_algorithms}
S.~G. Kobourov, \href{http://arxiv.org/abs/1201.3011}{Spring embedders and
  force directed graph drawing algorithms}, CoRR abs/1201.3011 (2012).
\newblock \href {http://arxiv.org/abs/1201.3011} {\path{arXiv:1201.3011}}.
\newline\urlprefix\url{http://arxiv.org/abs/1201.3011}

\bibitem{drawing_aesthetics_purchase}
H.~C. Purchase, R.~F. Cohen, M.~James, Validating graph drawing aesthetics, in:
  F.~J. Brandenburg (Ed.), Graph Drawing, Springer Berlin Heidelberg, Berlin,
  Heidelberg, 1996, pp. 435--446.

\bibitem{perceptual_organization_in_user_generated_grpah_layouts}
F.~van Ham, B.~Rogowitz, Perceptual organization in user-generated graph
  layouts, IEEE Transactions on Visualization and Computer Graphics 14~(6)
  (2008) 1333--1339.

\bibitem{shaefer-complexity-of-rectilinear-crossing-problem}
M.~Schaefer, Complexity of some geometric and topological problems, in:
  D.~Eppstein, E.~R. Gansner (Eds.), Graph Drawing, Springer Berlin Heidelberg,
  Berlin, Heidelberg, 2010, pp. 334--344.

\bibitem{geometric_heuristics_for_rectilinear_crossing_minimization}
M.~Radermacher, K.~Reichard, I.~Rutter, D.~Wagner, Geometric heuristics for
  rectilinear crossing minimization, ACM J. Exp. Algorithmics 24 (2019).

\bibitem{vertex_movement_radermacher_rutter}
M.~Radermacher, I.~Rutter, {Geometric Crossing-Minimization - A Scalable
  Randomized Approach}, in: M.~A. Bender, O.~Svensson, G.~Herman (Eds.), 27th
  Annual European Symposium on Algorithms (ESA 2019), Vol. 144 of LIPIcs,
  Schloss Dagstuhl--Leibniz-Zentrum fuer Informatik, Dagstuhl, Germany, 2019,
  pp. 76:1--76:16.

\bibitem{planar_Fary}
I.~Fáry, On straight-line representation of planar graphs, Acta Sci. Math.
  (1948) 229--233.

\bibitem{planarization_book}
C.~Gutwenger, P.~Mutzel, An experimental study of crossing minimization
  heuristics, in: G.~Liotta (Ed.), Graph Drawing, Springer Berlin Heidelberg,
  Berlin, Heidelberg, 2004, pp. 13--24.

\end{thebibliography}
\end{document}